\documentclass[a4paper,fleqn,usenatbib]{mnras}

\usepackage{graphicx} \usepackage{epsfig} \usepackage{natbib}
\usepackage{amsfonts} \usepackage{amsmath,amssymb,amstext}
\usepackage{float}

  \newcommand{\mstar}{M$_\ast$}
\newcommand{\lgmstar}{$\log_{10}($\mstar$/$M$_\odot)$}

\newcommand{\wrp}{$w_p(r_p)$} 
\newcommand{\mbh}{M$_{\mbox{BH}}$} \newcommand{\oiii}{[O {\sc iii}]}
\newcommand{\eddingtonrate}{$L$\oiii/\mbh}

\title[Stellar mass and color dependence of narrow-line AGN
  clustering] {Revisiting the clustering of narrow-line AGN in the
  local Universe: Joint dependence on stellar mass and color}

\author[Wang \& Li]{Lixin Wang\thanks{E-mail:
    wanglixin2016@mail.tsinghua.edu.cn} and Cheng Li\thanks{E-mail:
    cli2015@tsinghua.edu.cn} \\
Tsinghua center for Astrophysics and Physics Department, Tsinghua
University, Beijing 10084, China}

\date{Accepted ???. Received ???; in original form ???}

\begin{document}
\label{firstpage}
\pagerange{\pageref{firstpage}--\pageref{lastpage}} 
\maketitle
\begin{abstract}
  We investigate the clustering and dark halo properties for the
  narrow-line active galactic nuclei (AGN) in the SDSS,  particularly
  examining the joint dependence on galaxy mass and color.  AGN in
  galaxies with blue colors or massive red galaxies with $M_\ast\ga
  10^{10.5}M_\odot$ are found to show almost identical clustering
  amplitudes at all scales to control galaxies of the same mass, color
  and structural parameters. This suggests AGN activity in blue
  galaxies or massive red galaxies is regulated by internal processes,
  with no correlation with environment. The antibias of AGN at scales
  between $\sim100$kpc and a few Mpc, as found in \cite{Li-06a} for
  the AGN as a whole, is observed only for the AGN hosted by galaxies
  with red colors and relatively low masses ($<10^{10.5}M_\odot$).  A
  simple halo model in which AGN are preferentially found at dark halo
  centers can reproduce the observational results, but requiring a
  mass-dependent central fraction which is a factor of $\sim$4 higher
  than the fraction estimated from the SDSS group catalogue. The same
  group catalogue reveals that the host groups of AGN in red
  satellites tend to have lower halo masses than control galaxies,
  while the host groups of AGN in red centrals tend to form earlier,
  as indicated by a larger stellar mass gap between the two most
  massive galaxies in the groups.  Our result implies that the mass
  assembly history of dark halos may play an additional role in the
  AGN activity in low-mass red galaxies.
\end{abstract}
\begin{keywords}
Galaxies -- active galaxies -- clustering -- large scale of structure
\end{keywords}
\section{Introduction}
\label{sec:introduction}

Large redshift surveys of galaxies
accomplished in the past 1.5 decades have allowed the clustering of
optically selected AGN to be studied in great depth
\citep[e.g.][]{Porciani-Magliocchetti-Norberg-04,Wake-04,Croom-05,
  Constantin-06,Li-06a,Coil-07,Myers-07a,Shen-07,daAngela-08,Li-08,Ross-09,Shen-09a,
  Xu-12,Shen-13,Zhang-13,Shao-15,Chehade-16,Jiang-16,Shirasaki-18}.  For
instance, \citet{Wake-04} measured the two-point correlation (2PCF) of
narrow-line (type-2) AGN in the Sloan Digital Sky Survey
\citep[SDSS;][]{York-00}, finding the clustering amplitudes of AGN  to
be similar to that of luminous galaxies on scales between 0.2 to
$100h^{-1}$Mpc. By carefully matching the AGN sample with the control
sample of non-AGN in galaxy properties that are known to be correlated
with clustering (e.g. stellar mass, color, concentration and stellar
velocity dispersion), \citet{Li-06a} found the narrow-line AGN in the
SDSS to be more weakly clustered than control galaxies at intermediate
scales from 100kpc to a few Mpc, with no obvious difference on larger
scales. \citet{Jiang-16} compared the clustering of type-1 and type-2
AGN as well as normal galaxies from the SDSS, finding similar
clustering  amplitudes at scales larger than a few Mpc. \citet{Xu-12}
compared the clustering of galaxies with Narrow Line Seyfert 1 (NLS1)
and Broad Line Seyfert 1 (BLS1), also from the SDSS, and found no
significant difference at scales from a few tens of kpc to a few tens
of Mpc. These results demonstrate that optically-selected AGN are
hosted  by dark matter halos of similar masses to normal galaxies  of
similar properties, a conclusion that is true for different types of
AGN. The same conclusion was also obtained  by \citet{Pasquali-09}
using the SDSS galaxy group catalogue of \citet{Yang-07}.
The clustering of optical quasars from both 2dF Galaxy Redshift Survey 
\citep[2dFGRS;][]{Colless-01} and SDSS shows no or weak dependence
on luminosity at fixed redshift \citep[e.g.][]{Croom-05,daAngela-08,
Shen-09a,Chehade-16}, implying that QSOs of various luminosities
are hosted by dark matter halos of similar mass.

There has also been a rich history of studies on the clustering of AGN
detected in wavebands other than the optical
\citep[e.g.][]{Carrera-98, LaFranca-98, Basilakos-01, Francke-08,
  Plionis-08, Coil-09, Hickox-09, Mandelbaum-09, Basilakos-Plionis-10,
  Cappelluti-10, Donoso-10, Hickox-11, Starikova-11, Hickox-12,
  Krumpe-12, Mountrichas-Georgakakis-12, Koutoulidis-13, Worpel-13,
  Fanidakis-13, Donoso-14, Georgakakis-14, Allevato-14a, Allevato-14b,
  Krumpe-15, Mendez-16, Mountrichas-16, Koutoulidis-16, Allevato-16,
  Ballantyne-17, Magliocchetti-17, Retana-Montenegro-17, Aird-18, Hale-18,
  Krumpe-18, Plionis-18, Powell-18, Melnyk-17}.  When compared to
optically selected AGN samples, some authors found similar clustering
properties for X-ray AGN \citep[e.g.][]{Krumpe-12, Krumpe-15,
  Powell-18},  while many others found different clustering
amplitudes, bias factors and/or halo masses for a variety of X-ray
selected AGN samples.   X-ray selected AGN with high luminosities are
clustered more strongly than those with low luminosities
\citep{Krumpe-Miyaji-Coil-10, Cappelluti-10, Koutoulidis-13,
  Krumpe-15, Plionis-18}. Radio-loud quasars from SDSS are found in
denser environment than radio-quiet quasars \citep{Shen-09a}, while
most radio-loud AGN in SDSS are clustered more strongly than
radio-loud QSOs even when the AGN and QSO samples are matched in both
black hole mass  and radio luminosity \citep{Donoso-10}. Obsucred AGN
tend to reside in denser environments than unobscured AGN, as found
for both the AGN selected from mid-infrared colors from the WISE
survey \citep{Donoso-14} and those with hard X-ray detections from the
Swift/BAT survey \citep{Powell-18}. A related result is obtained in
\citet{Shao-15} where the AGN with higher mid-infrared luminosities
have more close companions than the AGN with lower mid-infrared
luminosities.

In this work, we would like to revisit the clustering of narrow-line
AGN in the SDSS, extending the earlier work of \citet{Li-06a}
(hereafter L06) by particularly examining the co-dependence on the
stellar mass and color of the host galaxies. In L06, the clustering
measurements and modelling were done for all the AGN as whole. We know
that, however, AGN are not a random subset of the general population
of galaxies \citep[e.g.][]{Kauffmann-03c}, and clustering depends
strongly  on galaxy properties, with stellar mass and color showing
the  strongest dependence \citep[e.g.][]{Li-06b}. It is interesting to
see how the AGN are clustered, particuarly on the intermediate scales
where the AGN antibias was observed, if divided into subsamples
according to both stellar mass and color of their host galaxies.  In
addition to measuring the clustering properties, we will also make use
of the SDSS galaxy group catalog constructed by \citet{Yang-07} from
the SDSS/DR7 galaxy sample.  This catalogue allows us to examine the
correlation of AGN with the central/satellite division, as well as the
properties of dark matter halos such as dark matter mass and mass
assembly history, as indicated by the luminosity or stellar mass gap
between the two most dominating galaxies in a given group.  

We're also interested to know whether the simple halo-based model
proposed in L06 can also successfully reproduce the co-dependence of
AGN clustering on stellar mass and colour.  In that model, AGN are
assumed to occur more often in the central galaxies of dark matter
halos, because a higher central fraction was shown to be able to
effectively reduce clustering amplitudes at scales below a few Mpc. In
fact, there have been observational evidence for the centers of
groups/clusters as preferred environment for both optical AGN and
radio-loud AGN \citep{Best-04, Croston-05, Mandelbaum-09,
  Pasquali-09}. For instance, using the SDSS group catalogue of
\cite{Yang-07}, \citet{Pasquali-09} find AGN activity to be suppressed
in satellite galaxies compared to central galaxies at fixed stellar
mass. In addition, the occurrence of optical satellite AGN depends
weakly on halo mass, and is not correlated with distance to the group
center. However, some other studies found lower fractions of AGN at
group/cluster centers. For instance, \citet{vonderLinden-10} found the
fraction of red galaxies with optical AGN decreases towards the group
center, while the occurrence of AGN in star-forming galaxies is
independent of clustercentric radius. \citet{Hwang-12} and
\citet{Manzer-DeRobertis-14} also found the fraction of AGN  to
decrease with decreasing the clustercentric radius.  A similar
conclusion is obtained for luminous X-ray AGN in massive  clusters
\citep{Haines-12}. 

The structure of the paper is as follows. In the next section we
describe the different samples and simulation data, as well as the
measurements of galaxy properties to be used in the next sections.  In
Section 3 we present our measurements of clustering for AGN samples
selected by mass and color, and compare the results with
carefully-matched control galaxy samples. In Section 4 we examine the
central fractions of AGN and the properties of their host groups using
the SDSS group catalogue. In Section 5 we then perform halo-based
modelling to interpret our observational results. We summarize our
work in the last section.

\section{Data}
\label{sec:data}

The data analyzed in this study are drawn from the Sloan Digital Sky
Survey data release 7 \citep[SDSS/DR7;][]{Abazajian-09}, from which we
have constructed i) a reference sample representing the general
population of galaxies in the local Universe, ii) a random sample
which has the same selection effects as the reference sample, iii) a
narrow-line AGN sample which is a subset of the reference galaxy
sample, and iv) control galaxy samples to be compared with the AGN
sample, also selected from the reference sample. For a given
(sub)sample of AGN or the corresponding control sample, we quantify
the clustering by estimating the projected two-point cross-correlation
function (2PCCF) with respect to the reference galaxy sample. In
addition, we make use of the SDSS group catalog constructed from
SDSS/DR7 by \cite{Yang-07} to examine the role of the
central/satellite classification in determining the AGN activity, as
well as the Millennium Simulation \citep[][]{Springel-05} to perform
halo-based modelling of the AGN clustering. We describe the samples
and simulation data in the rest of this section.

\subsection{The reference galaxy sample and random sample}
\label{sec:reference_sample}

The reference galaxy sample is a magnitude-limited catalog constructed
from the New York University Value Added Galaxy Catalog (NYU-VAGC)
{\tt sample dr72}, which is a catalog of low-redshift galaxies (mostly
below $z=0.3$) from the SDSS/DR7, described in detail in
\cite{Blanton-05c} and publicly available at
http://sdss.physics.nyu.edu/vagc/. The reference sample contains about
half a million galaxies with $r$-and Petrosian apparent magnitude of
$r<17.6$, $r$-band Petrosian absolute magnitude in the range
$-24<M_{^{0.1}r}<-16$, and spectroscopically measured redshift in the
range $0.01<z<0.2$. Here $M_{^{0.1}r}$ is corrected for evolution and
$K$-corrected to the value at $z=0.1$. 

The random sample is built up from the real galaxies in the reference
sample, following the method described in \cite{Li-06b}. In short, for
each real galaxy, we randomly generate 10 sky positions within the
mask of the reference sample and assign to each of them the redshift
and other properties (e.g. stellar mass and color) of the real
galaxy. This results in a unclustered sample, filling the same sky
area and having the same position- and redshift-dependent selection
effects as the real sample, but with 10 times larger sample
size. \cite{Li-06b} performed extensive tests, showing that random
samples constructed this way are valid for clustering measuring as
long as the survey area is substantially large and the effective
survey depth varies little across the survey area. Our reference
sample meets both requirements to good accuracy, covering $\ga 6000$
deg$^2$, complete down to $r=17.6$, and little affected by foreground
dust over the entire survey footprint.

\subsection{The AGN and control samples}
\label{sec:agn_sample}

The AGN sample is also selected from the reference sample. A galaxy in
the reference sample is identified as an AGN if it is classified as a
Seyfert or LINER on the BPT diagram
\citep[][]{Baldwin-Phillips-Terlevich-81}. In particular we use the
diagram of [O {\sc iii}]$\lambda5007$/H$\beta$ versus [N {\sc
    ii}]$\lambda6583$/H$\alpha$, adopting the dividing criterion of
\cite{Kauffmann-03a} for the identification. We take the flux and
error measurements of the relevant emission lines from the MPA/JHU
SDSS
catalog\footnote{http://www.mpa-garching.mpg.de/SDSS/}\citep{Tremonti-04,
  Brinchmann-04}. Following \cite{Brinchmann-04} we require all the
four emission lines ([O {\sc iii}]$\lambda5007$, H$\beta$, [N {\sc
    ii}]$\lambda6583$, H$\alpha$) to be significantly detected, each
with a signal-to-noise ratio S/N$>3$, in order for the galaxy to
appear on the BPT diagram. In addition, we have added a number of 52,781 
low-S/N AGN into our sample, which have S/N$>3$ only in H$\alpha$ and
[N {\sc ii}]$\lambda6583$ lines and [N {\sc
    ii}]$\lambda6583$/H$\alpha>0.6$, again following
\citet{Brinchmann-04}. Our sample consits of a total of 104,817 AGN.

In this work we will study the dependence of AGN clustering on the
properties of both the host galaxies and the AGN themselves. Galaxy
properties considered include stellar mass \mstar\ and optical color
$g-r$. For AGN we consider two parameters: black hole mass \mbh\ and
the accretion rate relative to the Eddington rate as quantified by
\eddingtonrate. We divide our AGN into subsamples according to these
properties, and for each subsample we construct a {\em control} sample
of galaxies selected from the reference sample, by simultaneously
matching five physical parameters: redshift, stellar mass, color,
concentration index ($C$) and central stellar velocity dispersion
($\sigma_\ast$). We apply the following matching tolerances:
${\Delta}cz < 500$km s$^{-1}$, ${\Delta}\log_{10}M_\ast < 0.1$,
${\Delta}(g-r) < 0.05$, ${\Delta}{\sigma}_{\ast} < 20$km s$^{-1}$,
${\Delta}C < 0.1 $. In some cases we use the 4000-\AA\ break
$D_n(4000)$ instead of $g-r$ with a tolerance of $D_n(4000)<0.05$.

\subsection{SDSS group catalog}
\label{sec:group_catalog}

The SDSS galaxy group catalog is constructed by \citet{Yang-07} by
applying a modified version of their halo-based group-finding
algorithm \citep{Yang-05c} to a sample of $\sim6.4\times10^5$ galaxies
from {\tt sample dr72} of the NYU-VAGC. We use the most massive galaxy
member as the central galaxy of each group. Accordingly, we classify
each galaxy in our reference sample and AGN sample as either a
``central galaxy'' or a ``satellite galaxy''. We will use this
classification to study the dependence of AGN clustering on the
central/satellite type of host galaxies.

\subsection{Physical parameters of AGN and galaxies}
\label{sec:physical_properties}

The measurements of galaxy and AGN properties as mentioned above are
taken from either the NYU-VAGC or the MPA/JHU SDSS database. We
briefly describe these parameters below and refer the reader to the
relevant papers for more detailed description.
\begin{description}
    \item[{\em Stellar mass:}] A stellar mass accompanies the NYU-VAGC
      release for each galaxy in our reference sample. This is
      estimated by \cite{Blanton-Roweis-07} based on the
      spectroscopically measured redshift and the SDSS Petrosian
      magnitude in the $u$, $g$, $r$, $i$, $z$ bands, assuming an
      initial mass function from \cite{Chabrier-03}. As described in
      Appendix of \cite{Guo-10} we have corrected the Petrosian mass
      to obtain a ``total mass'' using the SDSS model magnitudes.

    \item[{\em Optical color:}] The optical color $g-r$ is defined by
      the $g$-band and $r$-band Petrosian magnitudes, corrected for
      Galactic extinction and $K$-corrected to its value at $z=0.1$
      using the {\tt kcorrect} code \citep{Blanton-03b,
        Blanton-Roweis-07}. 

    \item[{\em $D_{n}(4000)$:}] This index is the amplitude of the
      4000-\AA\ break in the spectrum of each galaxy. We adopt the
      narrow version of the index defined in \cite{Balogh-99}. The
      $D_n(4000)$ index is sensitive to young stellar populations with
      age less than 1-2 Gyr, widely used as an indicator of the mean
      stellar age of galaxies. When compared to color, $D_n(4000)$ is
      less affected by dust attenuation.
    
    \item[{\em Concentration index:}] The concentration index is
      defined as $C=R_{90}/R_{50}$, the ratio of the radii enclosing
      90 and 50 per cent of the $r$-band light of the galaxy
      \cite[see][]{Stoughton-02}. 
    
    \item[{\em Central stellar velocity dispersion:}] The stellar
      velocity dispersion $\sigma_\ast$ is measured from the SDSS
      $3^{\prime\prime}$-fiber spectroscopy of each galaxy, and
      corrected for instrumental broadening. We have corrected the
      original $\sigma_\ast$ measurement provided in the MPA/JHU
      catalog to an aperture of $R_{50}/8$, where $R_{50}$ is the
      effect radius in $r$-band enclosing half of the total light of
      the galaxy. For this we adopt the relation found by
      \cite{Jorgensen-95}:
      $\sigma_{\ast,\mbox{corr}}/\sigma_{\ast,\mbox{fib}}=(8\times
      r_{\mbox{fib}}/R_{50})^{0.04}$, where
      $r_{\mbox{fib}}=1^{\prime\prime}.5$. 
    
    \item[{\em Black hole mass and Eddington ratio:}] For each AGN in
      our sample we estimate a black hole mass from the central
      $\sigma_\ast$ using the relation given in \cite{Tremaine-02}. We
      then estimate the black hole accretion rate relative to the
      Eddington rate by the ratio \eddingtonrate, in order to divide
      the AGN into subsmples of ``powerful'' and ``weak'' AGN. The
      \oiii\ luminosity is corrected for dust extinction using the
      flux ratio between $H\alpha$ and $H\beta$, following
      \citet{Wild-Heckman-Charlot-10}.
    
\end{description}

\subsection{Simulation data}
\label{sec:simulation}

We perform theoretical modelling to interpret our clustering
measurements based on the Millennium Simulation \citep{Springel-05}, a
large simulation of the $\Lambda$CDM cosmology with $10^{10}$
particles within a periodic box of size $L_{\mbox{box}}=500h^{-1}$Mpc
on a side, implying a particle mass of
$8.6\times10^8h^{-1}$M$_\odot$. Dark matter halos and subhalos at a
given output snapshot are identified using the {\tt SUBFIND} algorithm
described in \cite{Springel-01}, and merger trees are constructed to
describe the cosmic history of the growth of halos, which have formed
the basis for implementing empirical and semi-analytic models of
galaxy formation by many authors \citep[e.g.][]{Croton-06, Guo-11}. In
this study we will use the catalog of halo and subhalos at $z=0$, thus
focusing on the link of AGN/galaxies with dark matter (sub)halos in
the local Universe.

\section{Observational measurements of AGN clustering}
\label{sec:clustering}

\subsection{Clustering measure}
\label{sec:clustering_measure}

For a given AGN sample or its control galaxy sample ({\tt Sample Q}),
we measure the clustering by estimating the projected two-point
cross-correlation function (2PCCF) with respect to the reference
galaxy sample constructed above ({\tt Sample D}), over a wide range of
spatial scales from a few tens of kpc up to a few tens of Mpc. We
first estimate the 2PCCF in the redshift space, $\xi^{(s)}(r_p, \pi)$,
by
\begin{equation}
    \xi^{(s)}(r_p, \pi) =
    \frac{N_R}{N_D}\frac{QD(r_p,\pi)}{QR(r_p,\pi)}-1,
\end{equation}
where $N_D$ and $N_R$ are the number of galaxies in {\tt Sample D} and
in the random sample ({\tt Sample R}); $r_p$ and $\pi$ are the
separations perpendicular and parrallel to the line of sight;
$QD(r_p,\pi)$ and $QR(r_p, \pi)$ are the cross-pair counts between
{\tt Samples Q}, and {\tt D} and between {\tt Samples Q} and {\tt R},
respectively. 

The projected 2PCCF, $w_p(r_p)$, is then estimated by integrating
$\xi^{(s)}(r_p,\pi)$ over $\pi$:
\begin{equation}
    w_p(r_p) = \int_{\infty}^{\infty}\xi^{(s)}(r_p,\pi)d\pi =
    \Sigma_{i}\xi^{(s)}(r_p,\pi_i)\Delta\pi_i,
\end{equation}
where the summation runs from $\pi_1=-39.5h^{-1}$Mpc to
$\pi_{80}=39.5h^{-1}$Mpc, with $\Delta\pi_i$=1$h^{-1}$Mpc. We have
corrected the effect of fiber collisions following the method
described in L06. The errors on the \wrp\ measurements are estimated
using the bootstrap resampling technique
\citep{Barrow-Bhavsar-Sonoda-84, Mo-Jing-Boerner-92}.

\subsection{Clustering of the full AGN sample}
\label{sec:all_clustering}

\begin{figure}
\centering
\includegraphics[width=\columnwidth]{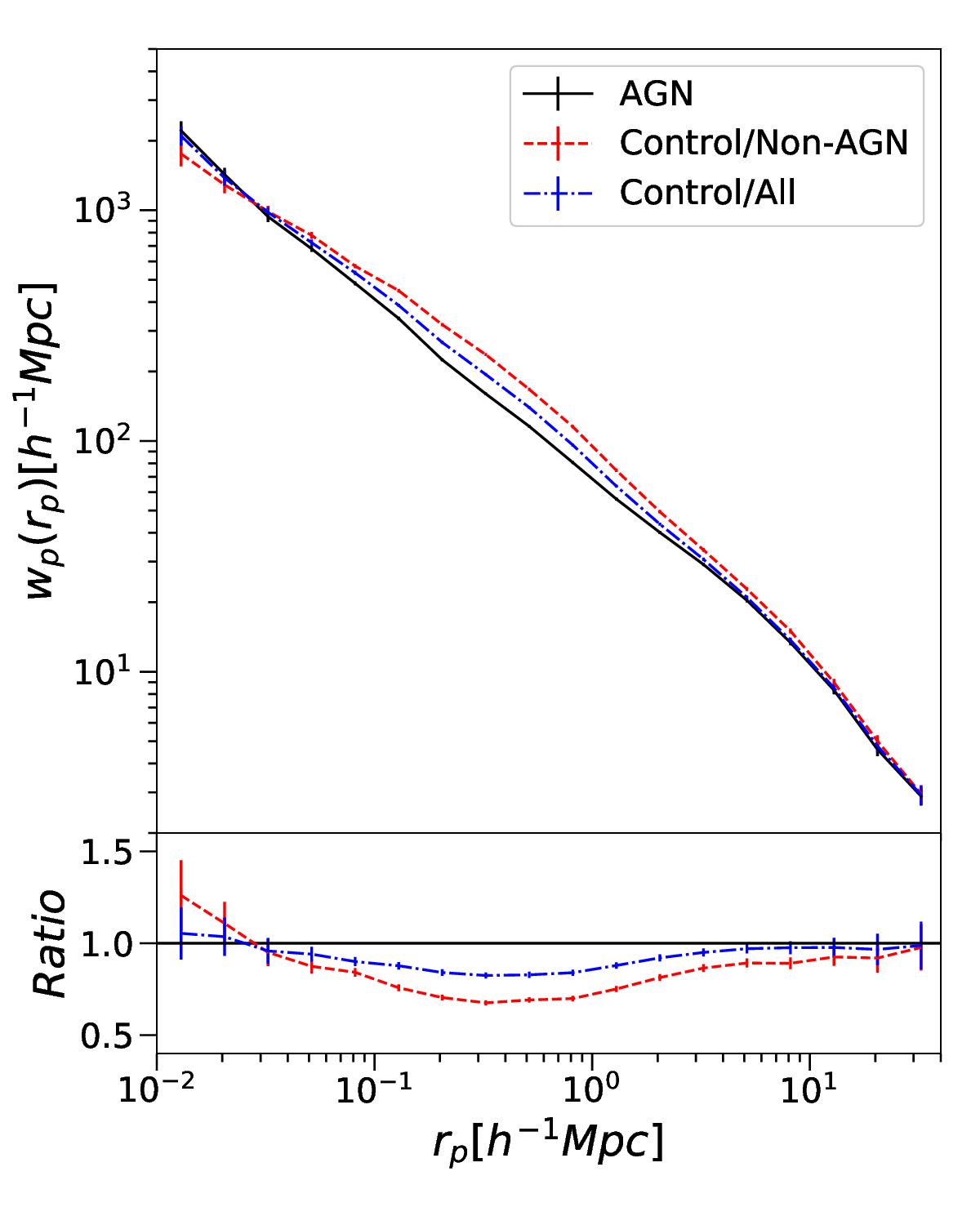}
\caption{Top: plotted as black crosses connected by the solid line is
  the projected cross-correlation function $w_p(r_p)$ between the full
  AGN sample and the reference galaxy sample.  The dashed red line and
  the dotted blue line present the $w_p(r_p)$ with respect to the same
  reference sample, as measured for two control samples, constructed
  from the galaxies with no AGN and the full reference sample,
  respectively.  Bottom: ratio of the $w_p(r_p)$ measurement of AGN to
  that of the control samples. Symbols/lines are the same as in the
  top panel.}
\label{fig:all_AGNs}
\end{figure}

We begin by estimating \wrp\ for the full AGN sample and the
corresponding control sample of galaxies, both with respect to the
reference sample. The measurements are shown in
Figure~\ref{fig:all_AGNs}. The upper panel displays the amplitudes of
the \wrp\ as a function of $r_p$, with the black line for the AGN
sample and the blue dotted line for the control sample. In the lower
panel the blue dotted line shows the ratio of the \wrp\ between the
AGN and the control sample, which is a measure of the scale-dependent
{\em bias} of the AGN relative to control galaxies of similar
properties.  The two samples are indistinguishable in clustering
amplitude on both large scales with $r_p$ above a few Mpc and small
scales with $r_p$ below a few tens of kpc. At intermediate scales the
AGN are clustered less strongly when compared to the control
galaxies. This {\em anti-bias} is strongest at a few $\times100
h^{-1}$kpc, and becomes weaker when one goes to both larger and
smaller scales. 

The AGN {\bf anti-bias} at intermediate scales was originally found in
L06. We note that the control samples in that paper were constructed
in a slightly different way, in the following two aspects. First, L06
considered only {\em non-AGN}, thus excluding the AGN from the control
samples. Therefore, the anti-bias discovered in that work was actually
the difference between galaxies with an AGN and those without an
AGN. Second, as described in Section~\ref{sec:data}, the control
sample used in the current work is closely matched with the AGN sample
in five parameters including redshift, stellar mass, concentration,
central stellar velocity dispersion and optical color, while L06
considered the first four parameters only. Here we additionally
include $g-r$ color, considering the known fact that galaxy clustering
depends on color over a wide range of scales even when the stellar
mass is limited to a narrow range \citep[e.g.][]{Li-06b}. 

For a better comparison with L06, we have also constructed control
samples of non-AGN, which are the galaxies from our reference sample
that are neither identified as AGN on the BPT diagram, nor selected as
low-S/N AGN. We have constructed two control samples of non-AGN: one
matched in redshift, stellar mass, concentration and central stellar
velocity dispersion, thus in exactly the same way as in L06,  and the
another matched additionally in $g-r$ as in this work. In
Figure~\ref{fig:all_AGNs} we show the \wrp\ measurement and the
AGN-to-control \wrp\ ratio obtained using the non-AGN control sample
that is matched in five parameters, as the red dashed line in both
panels. The results for the control sample matched in four parameters
are very similar, and are not shown in the figure for clarity. The
anti-bias is more pronounced when the control sample includes non-AGN
only, with a minimum \wrp\ ratio of $\sim0.7$ occurring also at a few
$\times100 h^{-1}$kpc. When compared to the result from the control
sample of {\em all} galaxies, the anti-bias extends to larger scales,
with a \wrp\ ratio of $\sim0.9$ even at scales exceeding 10 Mpc. At
smallest scales the \wrp\ ratio seems to suggest a slightly positive
bias, which should not be overemphasized, however, given the large
errors at these scales. All these results are in very good agreement
with L06 (see their Figure~3). 

We conclude that the anti-bias of AGN, as originally reported in Paper
I with respect to non-AGN, still exists but becomes weaker when the
bias is measured relative to the general population of galaxies
regardless of nuclear activity. The weaker anti-bias is naturally
expected, since the inclusion of some AGN into the control sample must
have lowered the intermediate-scale clustering to some degree,
thus reducing the difference between the AGN and the control
sample. In what follows we will use the control samples selected from
{\em all} galaxies, unless otherwise stated. 

\subsection{Joint dependence on mass, color and $D_n(4000)$}

\begin{figure*}
	\begin{center}
	  \epsfig{figure=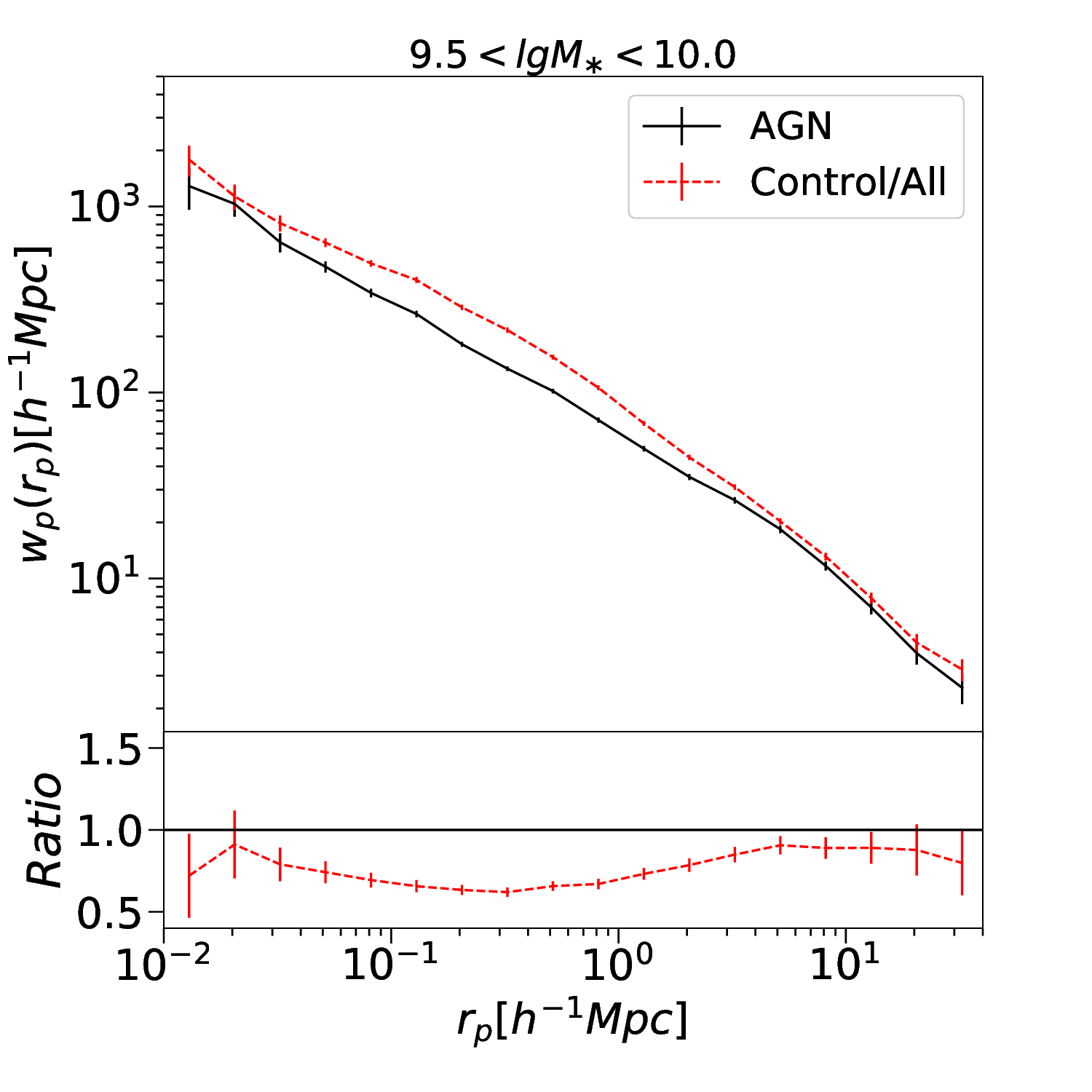,width=0.33\textwidth}
          \epsfig{figure=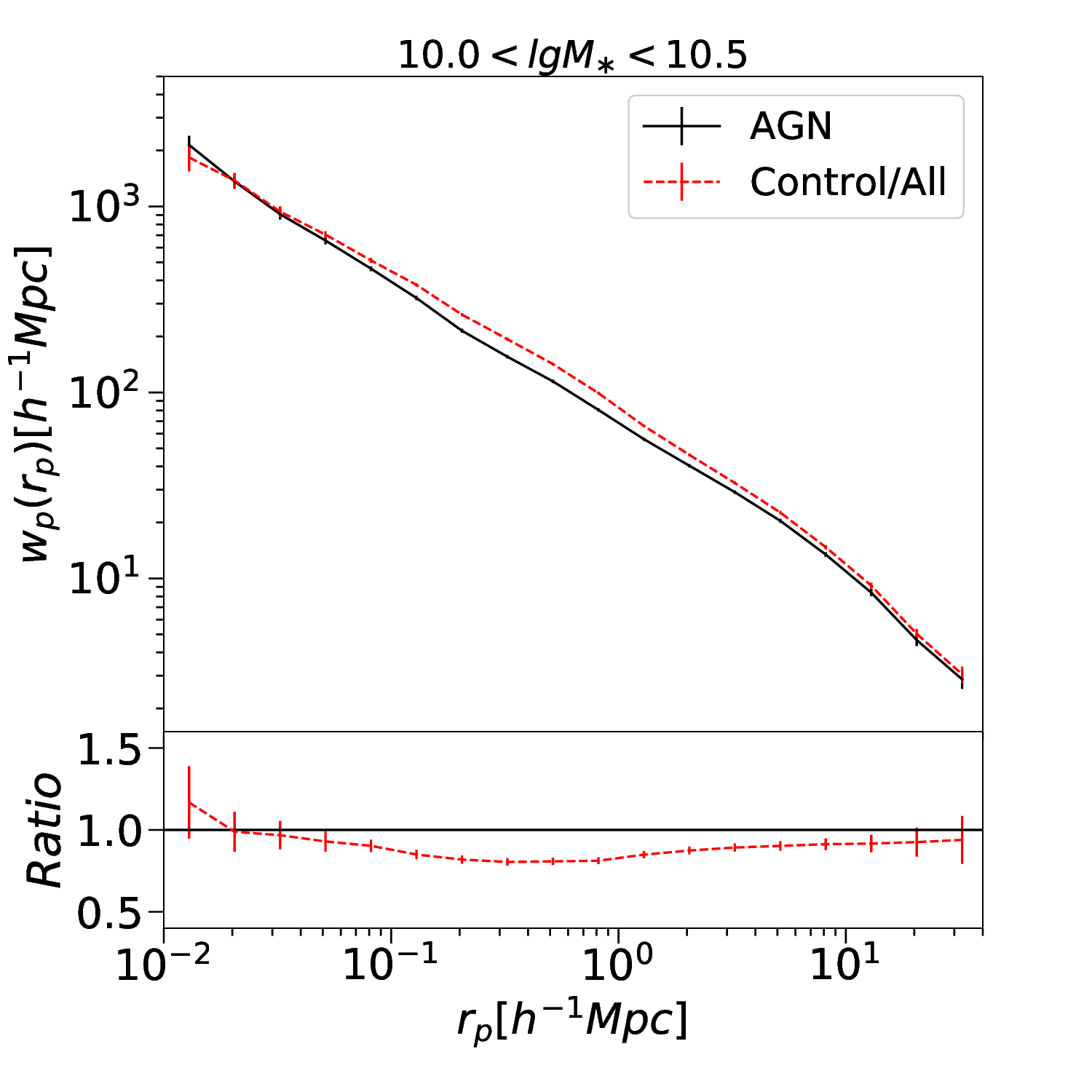,width=0.33\textwidth}
          \epsfig{figure=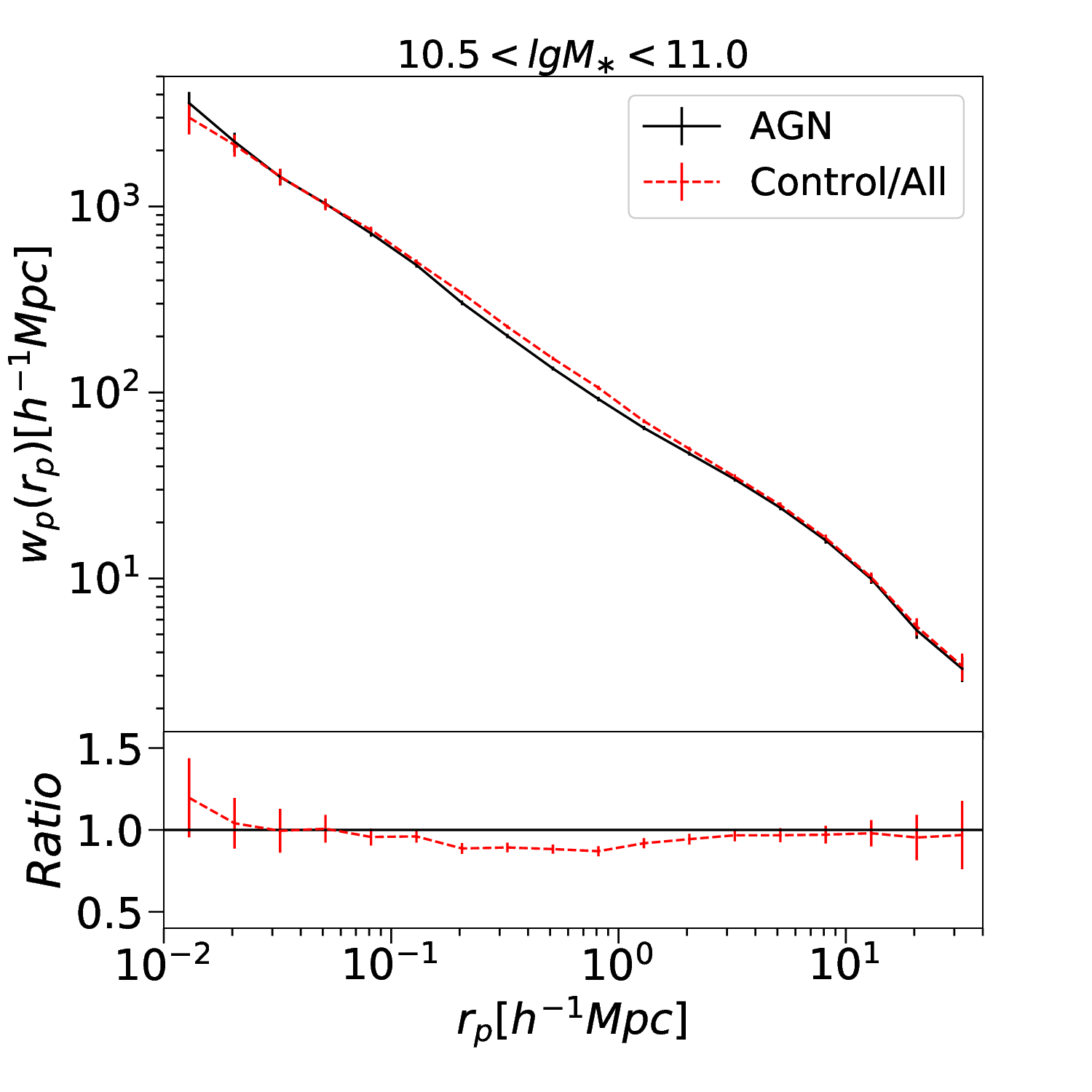,width=0.33\textwidth}
	\end{center}
	\caption{Top: $w_p(r_p)$ in different stellar mass intervals
          as indicated above each panel, for all AGN (red dashed
          lines) and for control samples (black solid lines). Bottom:
          ratio of $w_p(r_p)$ between the AGN and control samples in
          the top panel.}
	\label{fig:AGNs-stm}
\end{figure*}

\begin{figure*}
	\begin{center}
	  \epsfig{figure=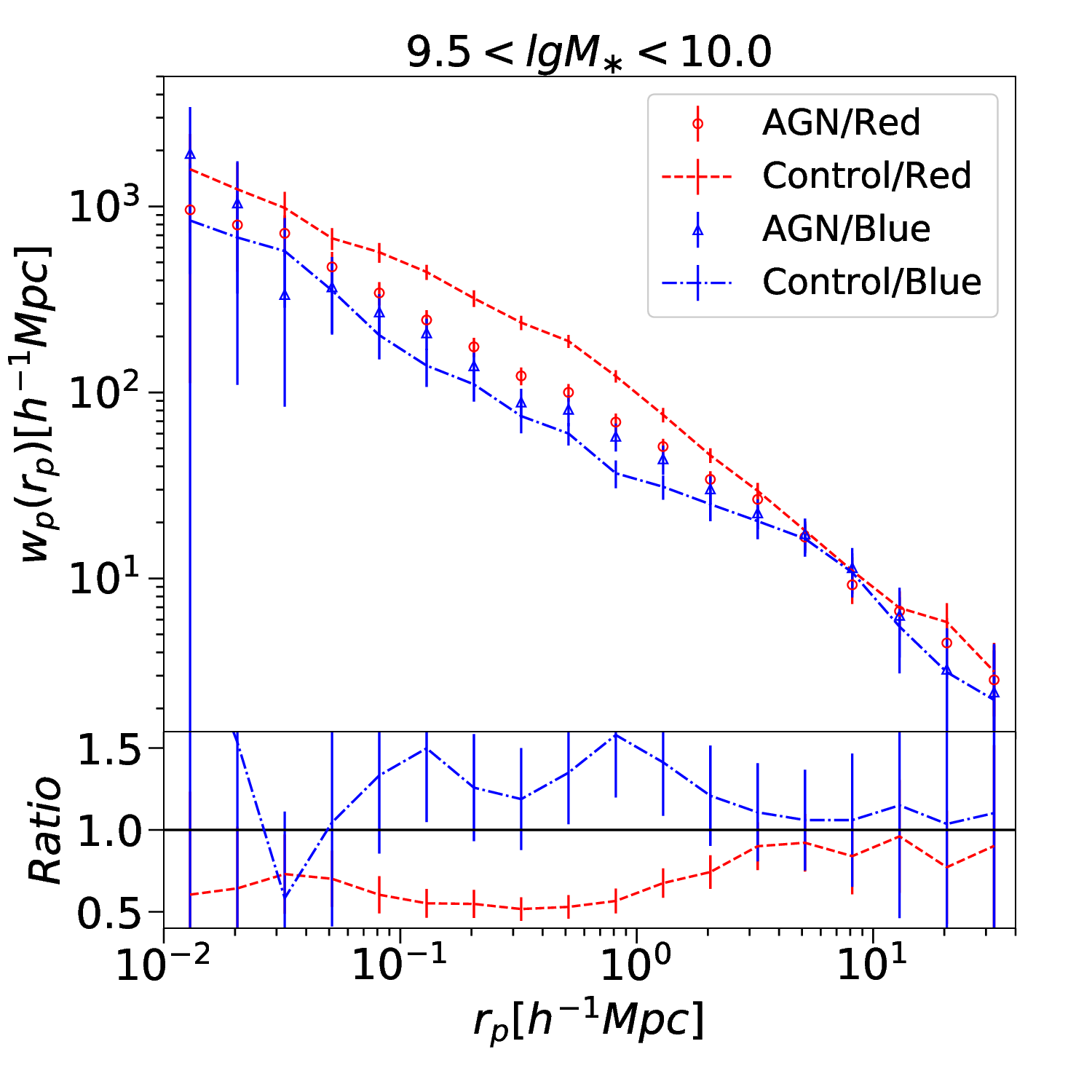,width=0.33\textwidth}
          \epsfig{figure=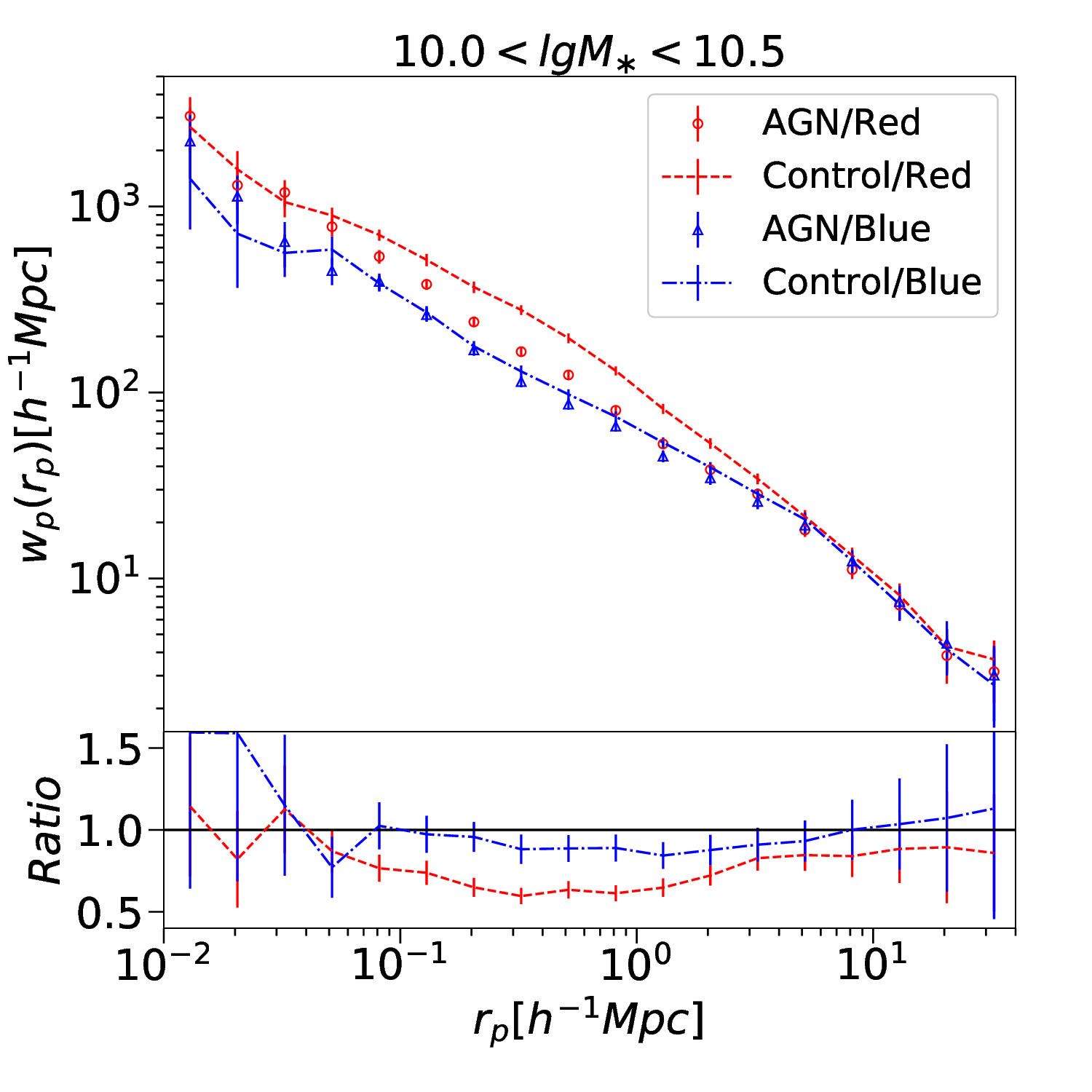,width=0.33\textwidth}
          \epsfig{figure=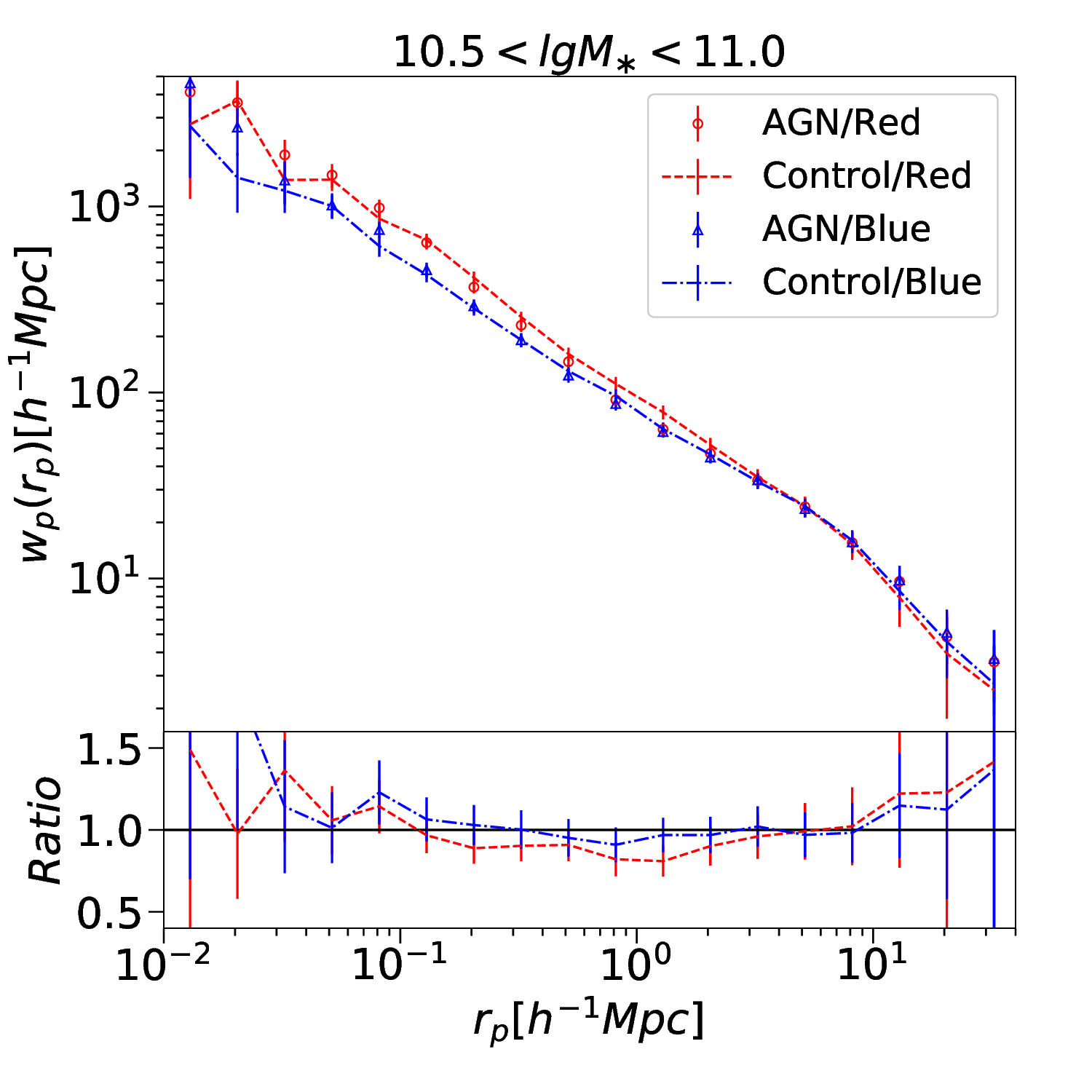,width=0.33\textwidth}
	\end{center}
	\caption{Top: $w_p(r_p)$ in different stellar mass intervals (
          indicated above each panel), for AGN hosted by red (red
          circles)  and blue (blue triangles) galaxies, and for the
          corresponding control samples. Bottom: ratio of $w_p(r_p)$
          between the AGN samples and their control
          samples. Symbols/lines are the same as in the top panel.}        
	\label{fig:agn-stm-color}
\end{figure*}

\begin{figure*}
	\begin{center}
	  \epsfig{figure=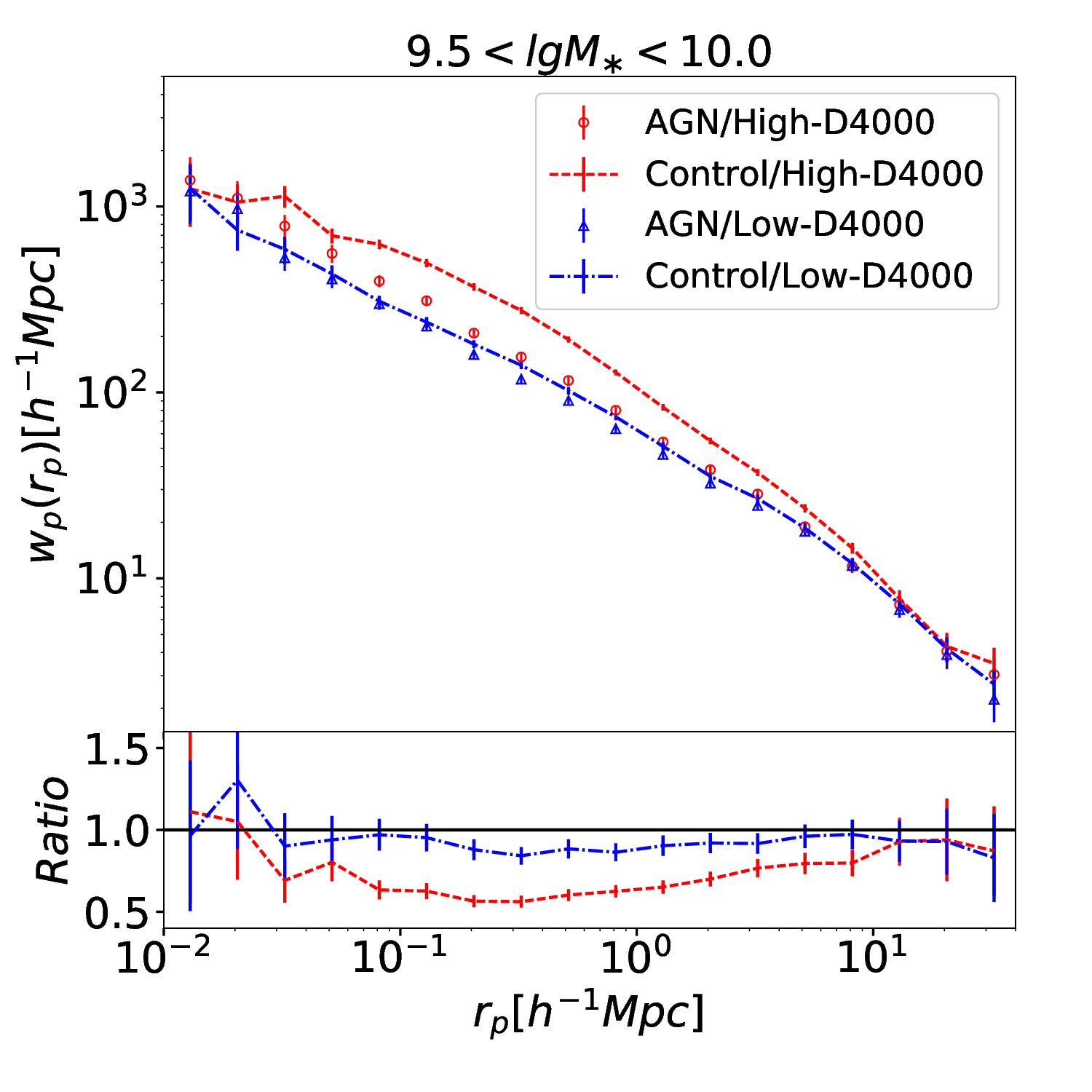,width=0.33\textwidth}
          \epsfig{figure=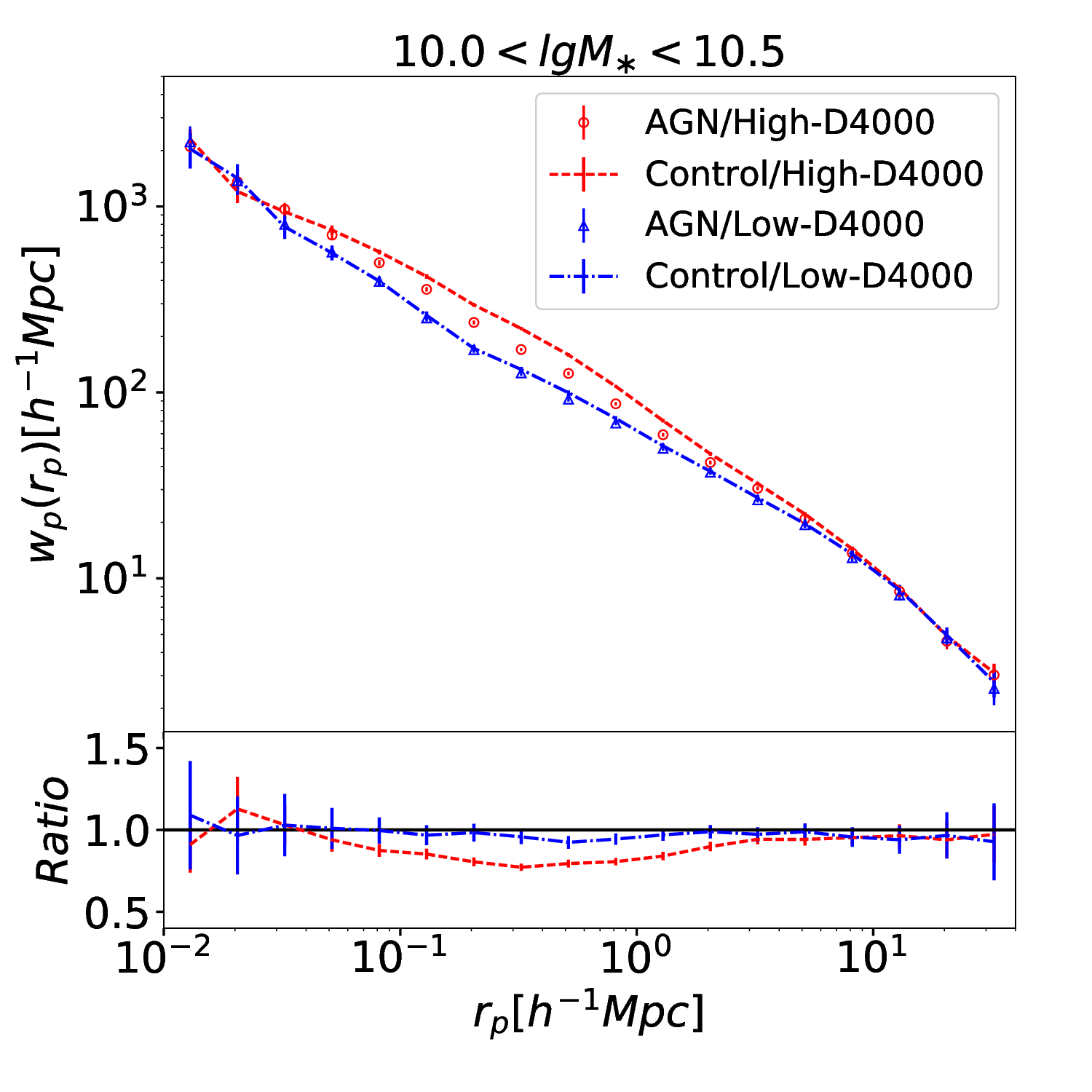,width=0.33\textwidth}
          \epsfig{figure=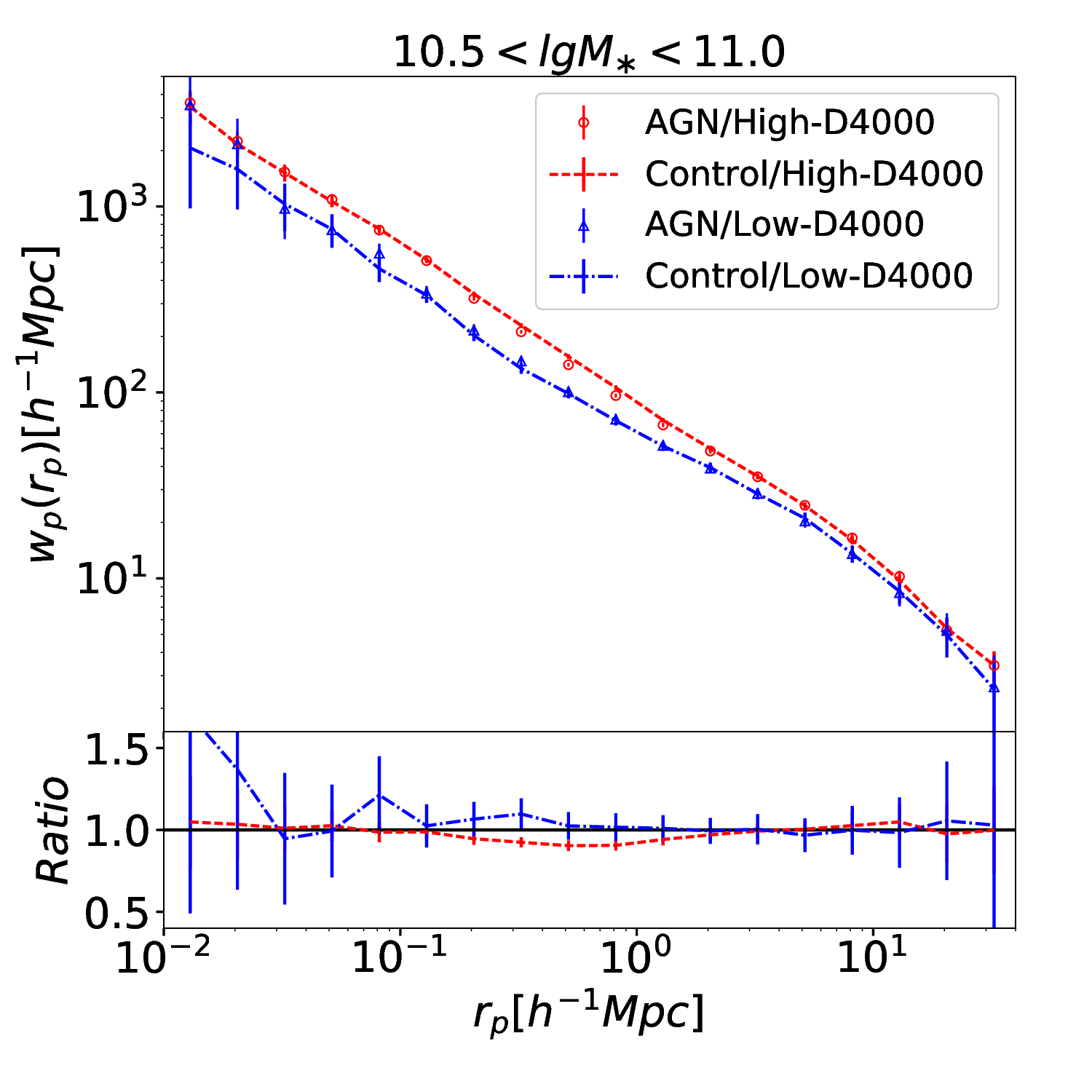,width=0.33\textwidth}
	\end{center}
	\caption{Top: $w_p(r_p)$ in different stellar mass intervals (
          indicated above each panel), for AGN hosted by galaxies with
          high $D_n(4000)$ (red circles) and low $D_n(4000)$ (blue
          triangles), and for the corresponding control
          samples. Bottom: ratio of $w_p(r_p)$ between the AGN samples
          and their control samples. Symbols/lines are the same as in
          the top panel.}        
	\label{fig:agn-stm-d4000}
\end{figure*}

Galaxy clustering is known to depend on a variety of properties. It is
interesting to study how the anti-bias of AGN depends on the
properties of their host galaxies. We consider three physical
parameters: stellar mass (\mstar), optical color ($g-r$) and the
4000-\AA\ break stength $D_n(4000)$. These are the galaxy properties
that are most related to environment \citep[e.g.][]{Kauffmann-04,
  Blanton-Moustakas-09} or clustering \citep[e.g.][]{Li-06b}. The
optical color and $D_n(4000)$ are similar in the sense that they are
indicator of the mean stellar age and recent star formation history of
galaxies.

We first divide all the AGN in our sample into three subsamples with
different stellar mass intervals: $9.5<$\lgmstar$<10$,
$10<$\lgmstar$<10.5$, and $10.5<$\lgmstar$<11$. For each AGN
subsample, we construct a control sample by selecting galaxies from
the reference sample, closely matched with the AGN sample in the four
parameters as described in Section~\ref{sec:data}. We then estimate
\wrp\ for the AGN subsamples and the corresponding control
samples. The \wrp\ measurements are shown in
Figure~\ref{fig:AGNs-stm}. Panels from left to right correspond to the
three stellar mass intervals. Panels in the upper row compare the
\wrp\ measurements for the AGN and control samples, while the lower
panels display the AGN-to-control \wrp\ ratio. 

As can be clearly seen from the figure, the anti-bias is most
remarkable for AGN that are hosted by galaxies with lowest stellar
masses, becoming less pronounced at higher stellar masses. For the
subsample with highest \mstar, the AGN and the control galaxies show
almost the same clustering behaviors, and the anti-bias at a few
$\times100 h^{-1}$kpc is no longer significantly seen. This result
shows that the anti-bias observed previously for the full AGN sample
is dominately contributed by the subset of AGN that are hosted by
relatively  low-mass galaxies. 

Next, for each stellar mass subsample, we further divide the AGN into
two subsets, with ``red'' or ``blue'' color, according to the $g-r$ of
their host galaxies. To the end, we have determined a mass-dependent
color divider: $(g-r)_{\mbox{cut}} = -1.399 + 0.2168$\lgmstar, based
on the distribution of the reference sample galaxies on the
\lgmstar\ versus $g-r$ plane. We have corrected the sample
incompleteness  caused by the volume effect of the survey by weighting
each galaxy by $1/V_{\mbox{max}}$, where $V_{\mbox{max}}$ is the
maximum volume over which the galaxy can be included in our sample. A
galaxy is classified as a red galaxy if its $g-r$ is larger than the
$(g-r)_{\mbox{cut}}$ at its stellar mass. Otherwise it is a blue
galaxy. For each AGN subsample we construct a control sample in the
same way as above. 

Figure~\ref{fig:agn-stm-color} displays the \wrp\ measurements for the
AGN subsamples selected by \mstar\ and color. Panels from left to
right are for the three mass intervals. In each panel, results for the
AGN in red (blue) hosts and the corresponding control sample are
plotted in red (blue) symbols and the dashed-red (dotted-blue)
line. We find that, when limited to blue galaxies, the AGN show almost
no or very weak anti-bias, and this is true at all scales and for all
masses. It is interesting that the anti-bias at intermediate scales is
hold only for AGN in red galaxies. This result clearly shows that the
overall anti-bias seen for the full AGN sample is dominated by those
AGN in low-mass red galaxies, while the AGN hosted by blue galaxies of
all masses or red galaxies of high masses are clustered in the same
way as the control galaxies.

Finally, we examine the co-dependence of AGN clustering on mass and
$D_n(4000)$, repeating the above analysis but using $D_n(4000)$
instead of $g-r$ to divide galaxies in a given mass range into
subsamples of high or low $D_n(4000)$.  Figure~\ref{fig:agn-stm-d4000}
shows the results which are very similar to what is seen from the
previous figure on the  co-dependence on mass and color. On one hand,
there is little difference in \wrp\ between the AGN of low $D_n(4000)$
and the control galaxies.  On the other hand, strong anti-bias is
observed for the AGN hosted  by galaxies of high $D_n(4000)$ and
intermediate-to-low masses.  The similarity is well expected, because
$g-r$ and $D_n(4000)$ are both related to the recent star formation
history of galaxies, as mentioned above. 

A common result from
Figures~\ref{fig:all_AGNs}-\ref{fig:agn-stm-d4000} is that, on large
scales with $r_p$ above a few Mpc, the AGN always show the same
clustering amplitudes as the control galaxies, and this is true for
all the mass intervals and for all the subsets selected by $g-r$ or
$D_n(4000)$. It is known that the amplitude of clustering at scales
larger than a few Mpc is an indicator of the dark matter halo mass of
galaxies. Therefore, our finding implies that the AGN and control
galaxies, once matched closely in the main properties, intend to
populate dark matter halos of similar mass. In addition, we also note
from both Figure~\ref{fig:agn-stm-color} and
Figure~\ref{fig:agn-stm-d4000} that, at fixed stellar mass and at the
intermediate scales, the AGN with red colors or high $D_n(4000)$ are
more clustered than the AGN with blue colors or low $D_n(4000)$,
although the red or high-$D_n(4000)$ AGN are anti-biased at these
scales relative to the control sample. 

\subsection{Dependence on black hole mass, AGN power and AGN type}

\begin{figure*}
	\begin{center}
	  \epsfig{figure=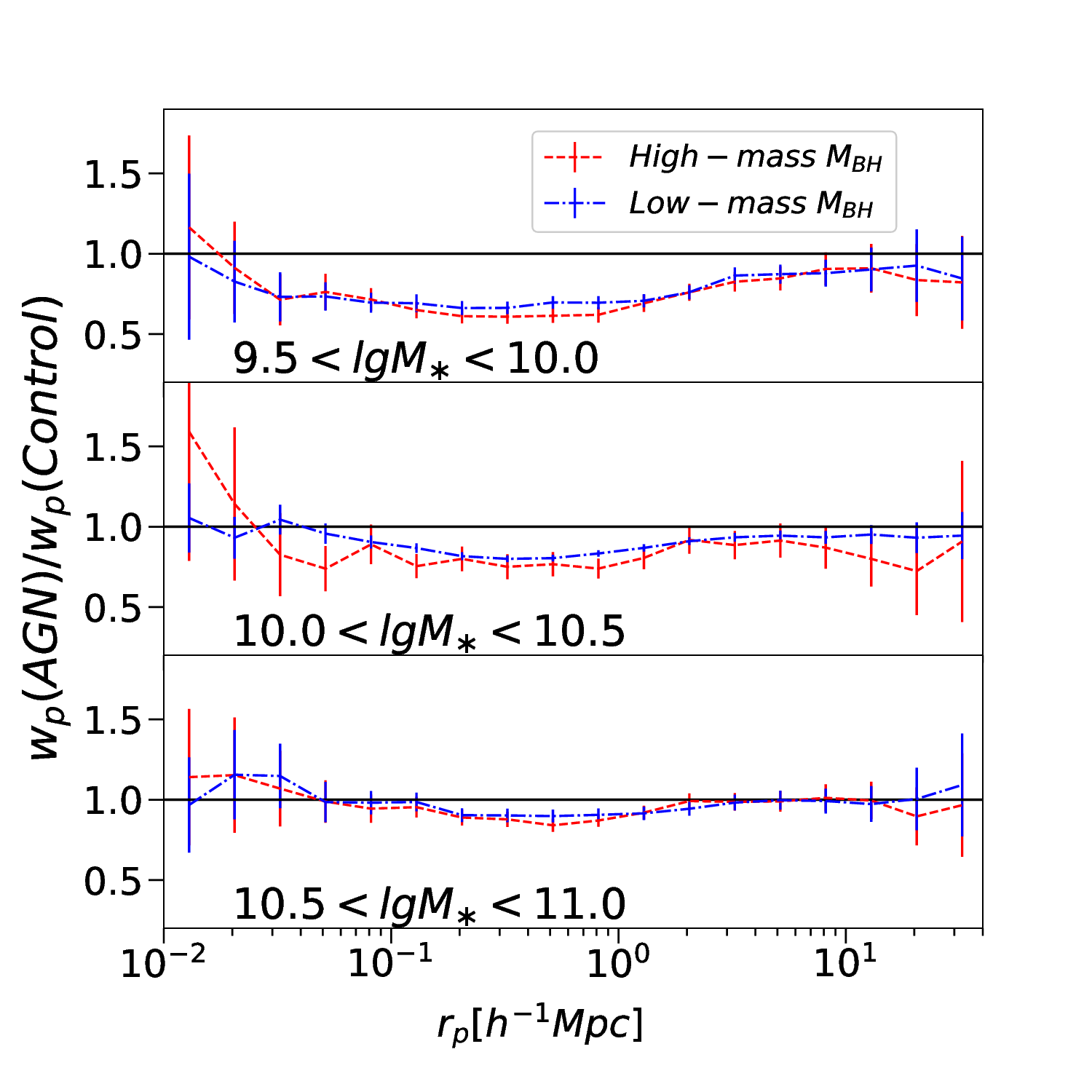,width=0.49\textwidth}
          \epsfig{figure=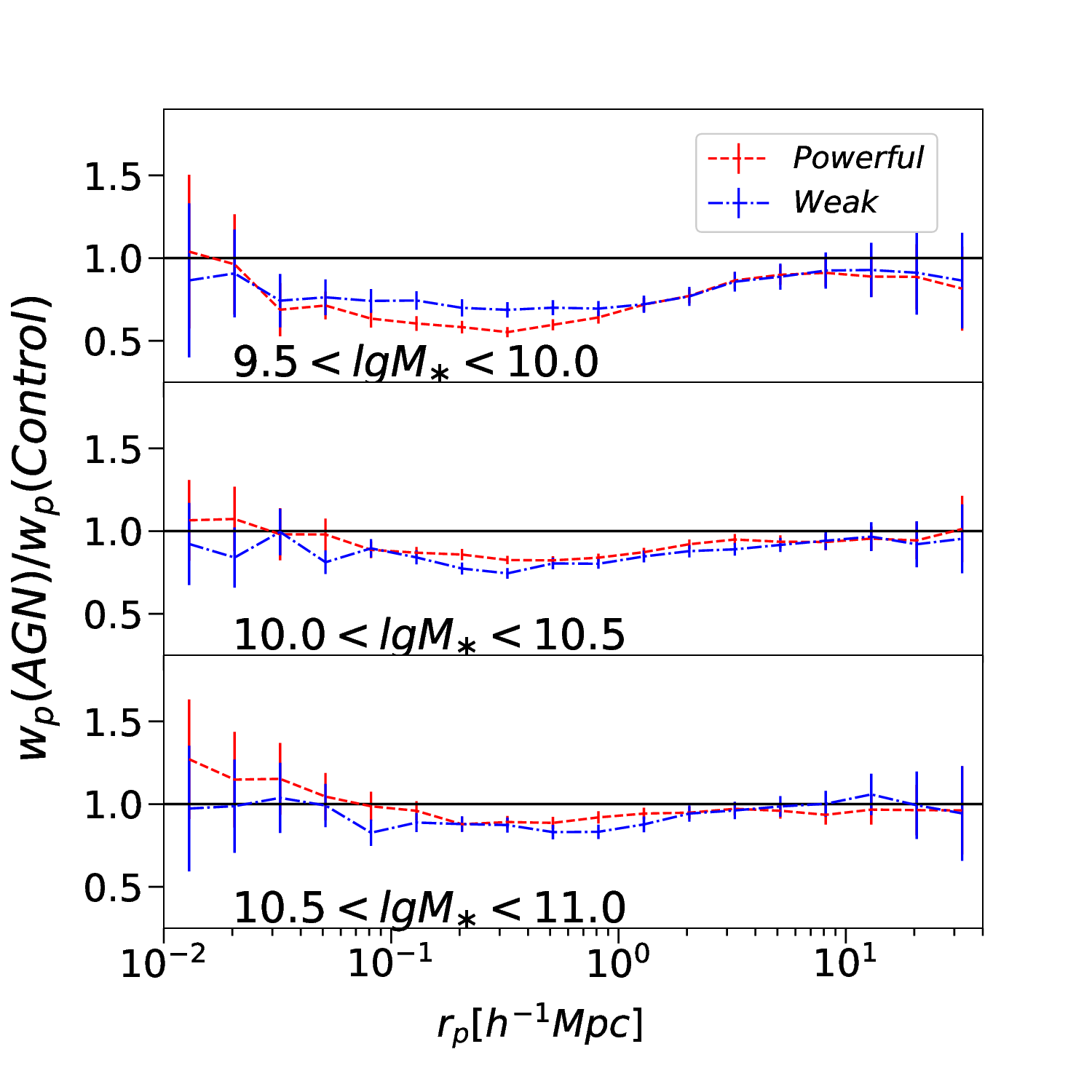,width=0.49\textwidth}
	\end{center}
	\caption{AGN-to-control ratio of $w_p(r_p)$ for galaxies in
          different stellar mass bins (indicated in each panel), and
          for AGN with high or low black mass (left panels), or AGN
          with high or low Eddington ratio (right panels). See the text
          for details.}
	\label{fig:agn-stm-bhtype}
\end{figure*}

\begin{figure*}
	\begin{center}
	  \epsfig{figure=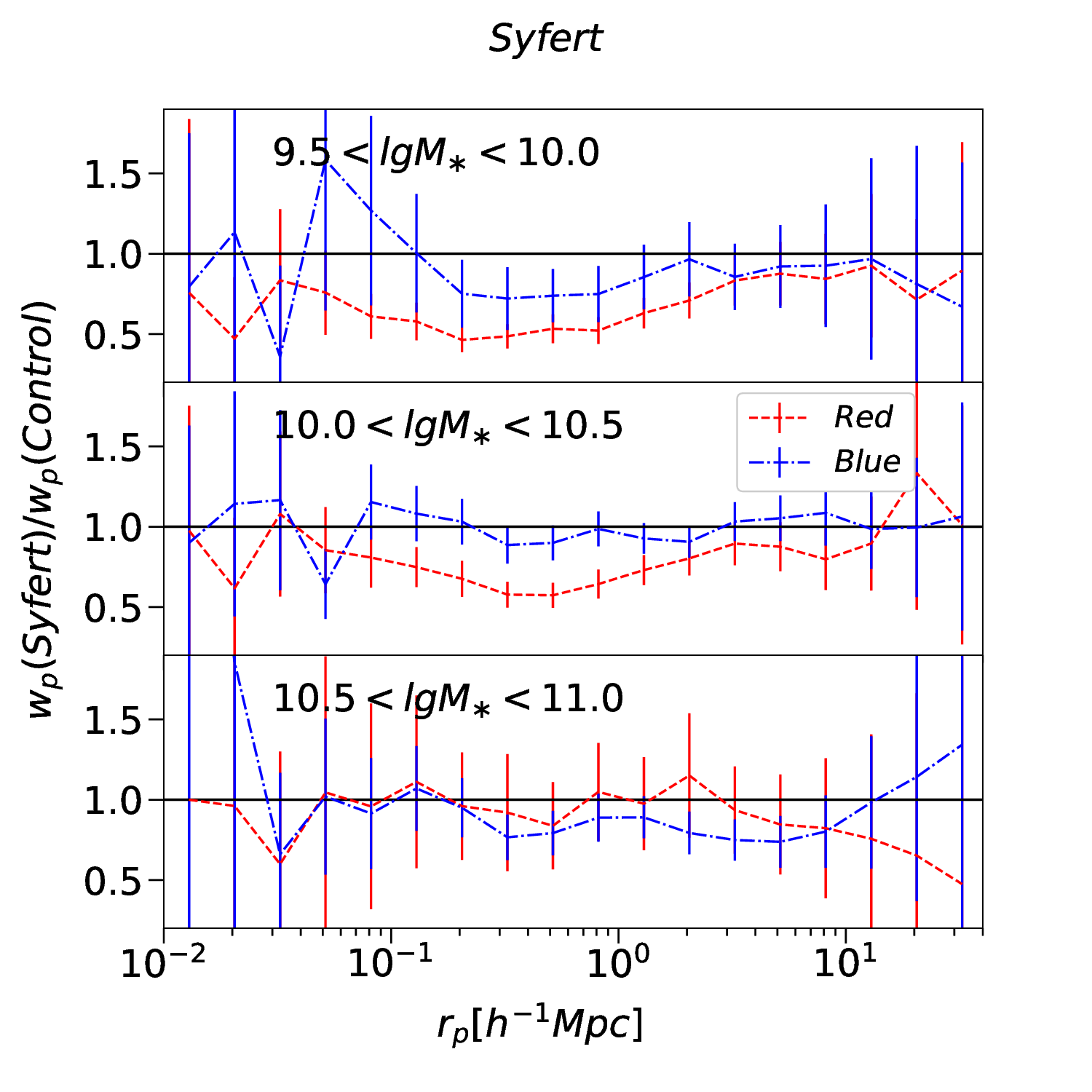,width=0.49\textwidth}
          \epsfig{figure=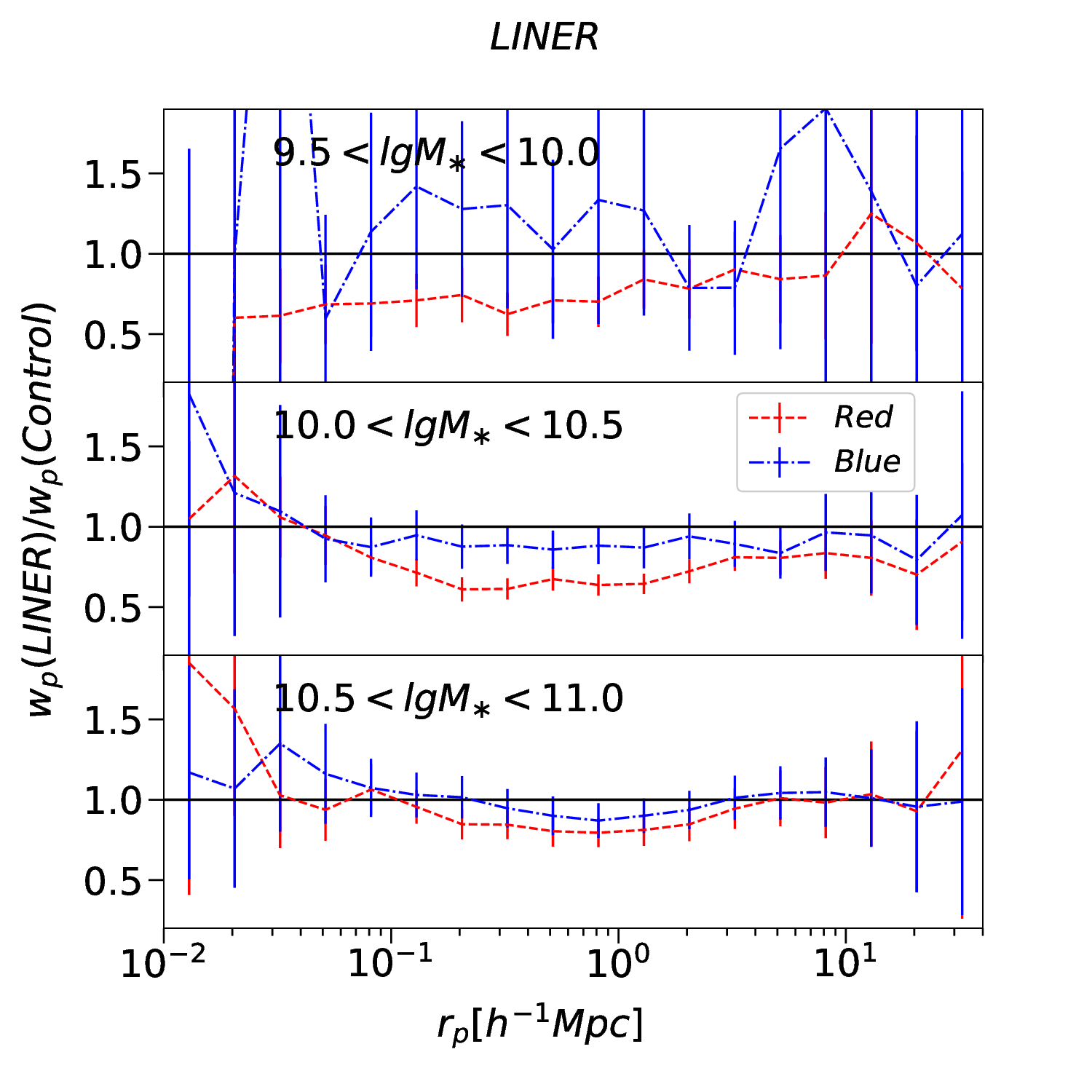,width=0.49\textwidth}
	\end{center}
	\caption{AGN-to-control ratio of $w_p(r_p)$ for different
          stellar  mass bins (indicated in each panel), and for
          Seyferts (left panels) and LINERs (right panels). In each
          panel, the red/blue lines are for AGN hosted by galaxies of
          red/blue colors.}
	\label{fig:AGNtype-stm-color}
\end{figure*}

In this subsection we further study the dependence of AGN clustering,
particularly the anti-bias at intermediate scales, on the properties
of the AGN themselves. As described in Section~\ref{sec:data}, we have
estimated a black hole mass \mbh\ and the accretion rate relative to
the Eddington rate \eddingtonrate\ for each AGN in our sample. For a
given stellar mass interval, we divide the AGN into two subsets with
either high or low \mbh, or with either high or low \eddingtonrate. We
adopt the median values of \mbh\ and \eddingtonrate\  in a given
stellar mass bin as the dividers, so that the two subsets have the
same sample size. A control sample is constructed for each subsample
in the same way as above. Figure~\ref{fig:agn-stm-bhtype} displays the
\wrp\ ratio between the AGN and control samples. In the left-hand
panel, the red and blue lines in each panel are results for the AGN
subsamples with high and low \mbh, respectively, while the three
panels are for the three stellar mass bins. Similarly, the right-hand
panel compares the AGN-to-control \wrp\ ratio for the ``powerful'' AGN
(high \eddingtonrate, red line) and the ``weak'' AGN (low
\eddingtonrate, blue line). 

As can be seen from Figure~\ref{fig:agn-stm-bhtype}, the clustering
amplitude shows very weak or no dependence on either \mbh\ or
\eddingtonrate, and this is true for all stellar masses and at all
scales probed. Slight difference is observed in the lowest mass range
($9.5<$\lgmstar$<10$, the top-right panel), where the powerful AGN
appear to show stronger anti-bias than the weak AGN on scales
$0.1h^{-1}$Mpc$<r_p<1h^{-1}$Mpc. This result is in broad agreement
with L06 which found a marginal tendency for powerful AGN to be more
strongly anti-biased than weak AGN at the highest Eddington
ratios. L06 also found the anti-bias to slightly depend on black hole
mass with more massive black boles showing stronger anti-bias.  This
effect may be attributed to the different way of constructing control
samples as we point out above.  However, the dependence on black hole
mass was rather weak and wasn't of high significance due to the large
error bars.

In Figure~\ref{fig:AGNtype-stm-color} we examine the co-dependence of
clustering on mass and color but separately for Seyferts (left panel)
and LINERs (right panel). In each case we show the AGN-to-control
\wrp\ ratios, compared for red and blue galaxies for a given stellar
mass range. Although the measurements become noisy due to smaller
sample sizes when compared to the measurements in previous figures,
the \wrp\ ratios of both Seyferts and LINERs depend on mass and color
in exactly the same way as the \wrp\ ratios of the whole AGN sample.
The intermediate-scale anti-bias is observed only for Seyferts/LINERs
hosted by low-mass red galaxies, and there is no effect for blue
galaxies or red galaxies of high masses. This finding shows that the
anti-bias at intermediate scales and its co-dependence on mass and
color are a general property of different types of AGN.

\section{Central fractions and halo properties based on the SDSS group catalog}

In L06, the antibias observed for the full AGN sample was  interpreted
by a higher fraction of central galaxies in the  AGN sample than the
control sample. A simple halo-based model  in which the central
fraction of AGN ($f_c$) is the only free parameter could well
reproduce the $w_p(r_p)$ measurement of both AGN and control
galaxies. In this section we make use of the SDSS galaxy group
catalogue (see \S\ref{sec:group_catalog}) to estimate $f_c$ for both
the AGN samples and the control samples, as a function of galaxy mass
and color. In addition,  we also examine the dark matter halo mass and
assembly history (indicated by the stellar mass gap between the most
massive and the second most massive galaxy in a  group), using the
same group catalogue. These parameters are known to be correlated with
clustering at different scales.

\subsection{Central fraction of AGN as a function of stellar mass and color}

\begin{figure}
  \begin{center}
    \epsfig{figure=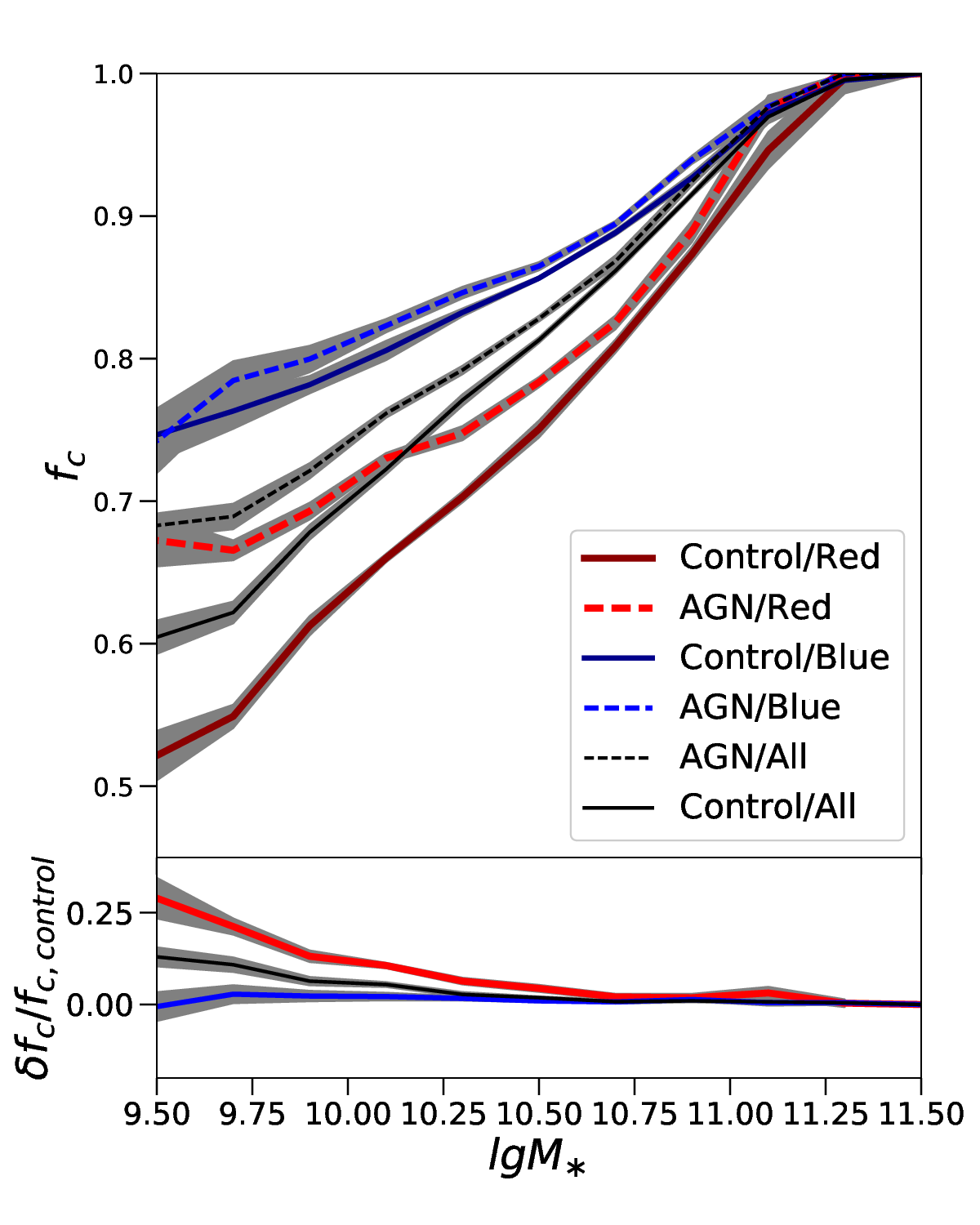,width=\columnwidth}
  \end{center}
  \caption{Top: central fraction $f_c$ is plotted as a function of
    stellar mass for all AGN (black dashed line), and for AGN hosted
    by red (red dashed line) and blue (blue dashed line) galaxies,
    compared to the $f_c$ of the corresponding control samples
    (black/red/blue solid lines). Bottom: relative difference of $f_c$
    between AGN and control samples. Lines are the same as in the top
    panel. In both panels, the lines are the median $f_c$ in given
    sample, and the shaded regions indicate the 1$\sigma$ scatter
    around the median.}
  \label{figs:agncolorfra}
\end{figure}

We aim to estimate the central fraction for AGN/control samples with
both red and blue colors, and in different stellar mass bins.  For
this purpose, we have selected a set of four volume-limited samples on
the plane of $\log_{10}(M_\ast)$ versus redshift out of the reference
sample, adopting the mass thresholds and redshift intervals as in
\cite{vonderLinden-10} (see their Fig.5).  The selection criteria
ensure that, for each of the four samples,  both the red and blue
populations are complete down to the chosen  mass threshold and within
the corresponding redshift range. We then match the samples with the
SDSS group catalog to obtain their central/satellite classification
and dark matter halo properties.

Figure~\ref{figs:agncolorfra} shows $f_c$ as a function of
$\log_{10}(M_\ast)$ for the full AGN sample and the AGN hosted by red and
blue galaxies. Results of the corresponding control samples are also
plotted for comparison. The relative difference of $f_c$ between the
AGN to control samples are plotted in the smaller panel of the same
figure.  Overall, $f_c$ increases with increasing stellar mass in all
the cases, reaching $100\%$ at the highest masses ($\log_{10}
M_\ast>11.25$).  At lower masses, $f_c$ at fixed mass differ from
sample to sample, but basically the blue populations including both
AGN and control galaxies have higher $f_c$ when compared to the red
populations at the same mass. 

Comparing the AGN and control samples, we find the $f_c$ of the full
AGN sample to be higher than that of the control sample at stellar
masses below $\sim10^{10.5}M_\odot$. The relative difference is
largest at lowest masses, $\sim13\%$ at $\log_{10}M_\ast\sim9.5$, and
decreases with mass. At masses above $\sim10^{10.5}M_\odot$ the AGN
and control samples present the same central fractions. More
interestingly, when the AGN are divided into subsets with red and blue
colors,  we find the same mass-dependent differences to be hold in the
red samples, and the AGN and control galaxies of blue colors present
very similar $f_c$ at all masses. We note that consistent results were
found in an earlier paper by \citet{Pasquali-09}.   

\subsection{Halo properties of AGN according to central/satellite division}

\begin{figure*}
  \begin{center}
    \epsfig{figure=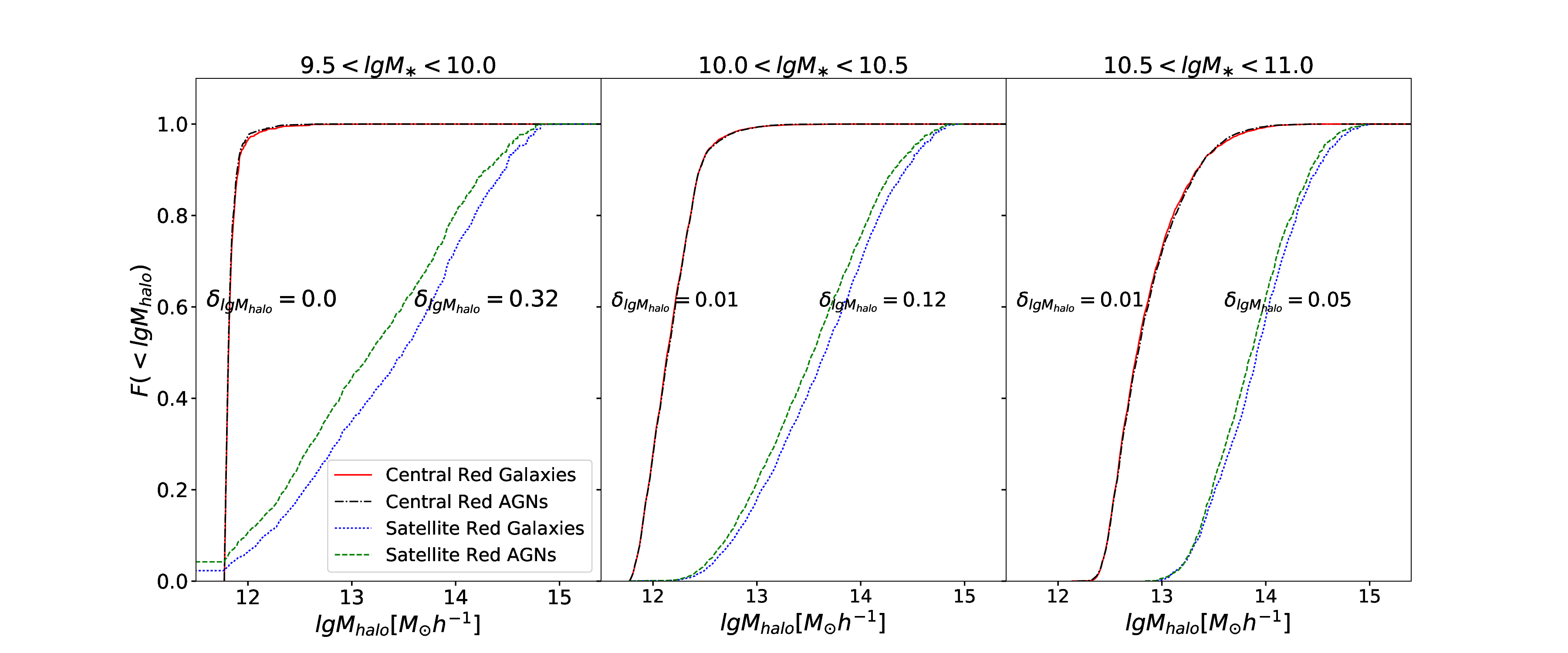,width=\textwidth}
    \epsfig{figure=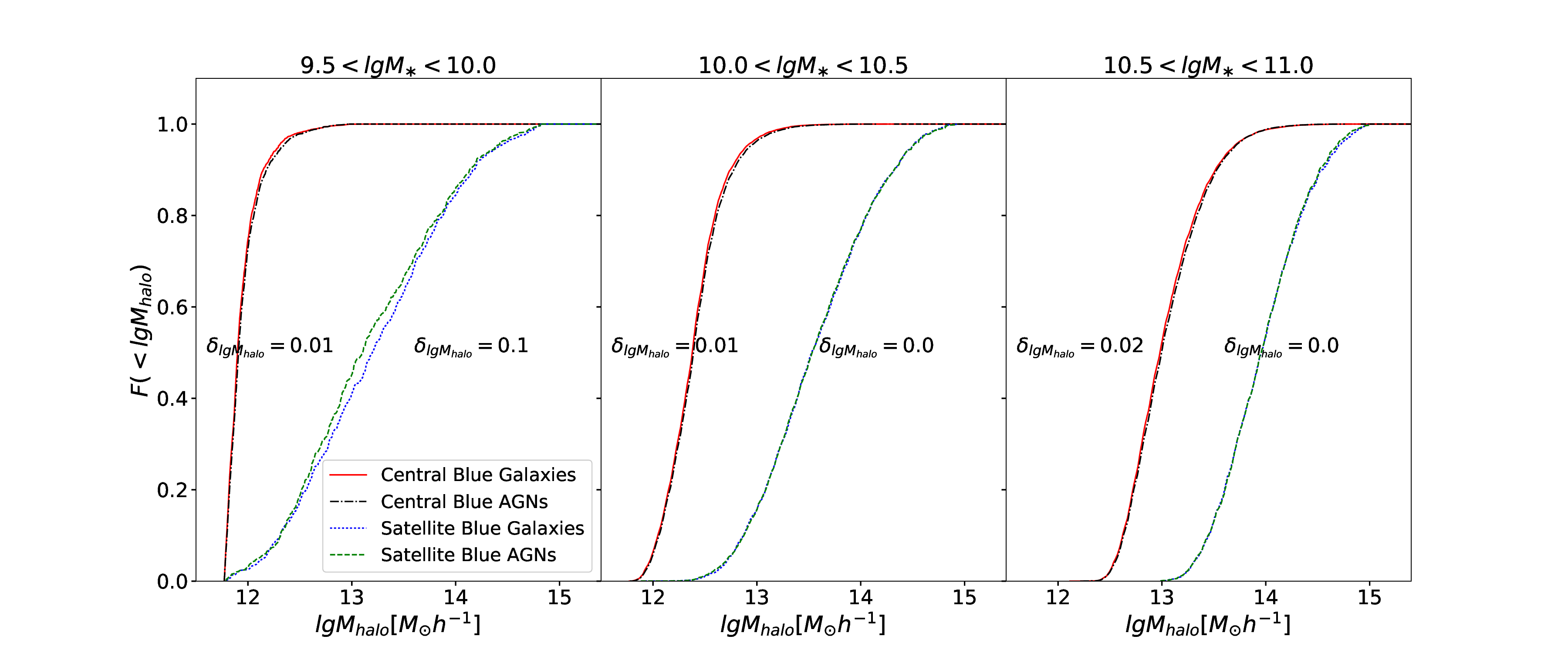,width=\textwidth}
  \end{center}
  \caption{Cumulative fractions of dark halo mass are plotted for AGN
    (dashed lines) and control galaxies (solid lines), in different
    stellar mass bins (indicated above each panel) and for red (upper
    panels) and blue (lower panels) colors. The difference in the
    median value of $\log_{10}(M_{halo})$ between each AGN sample and the
    corresponding control sample is indicated.}
  \label{fig:com-agn-gal-halo}
\end{figure*}

\begin{figure*}
   \begin{center}
      \epsfig{figure=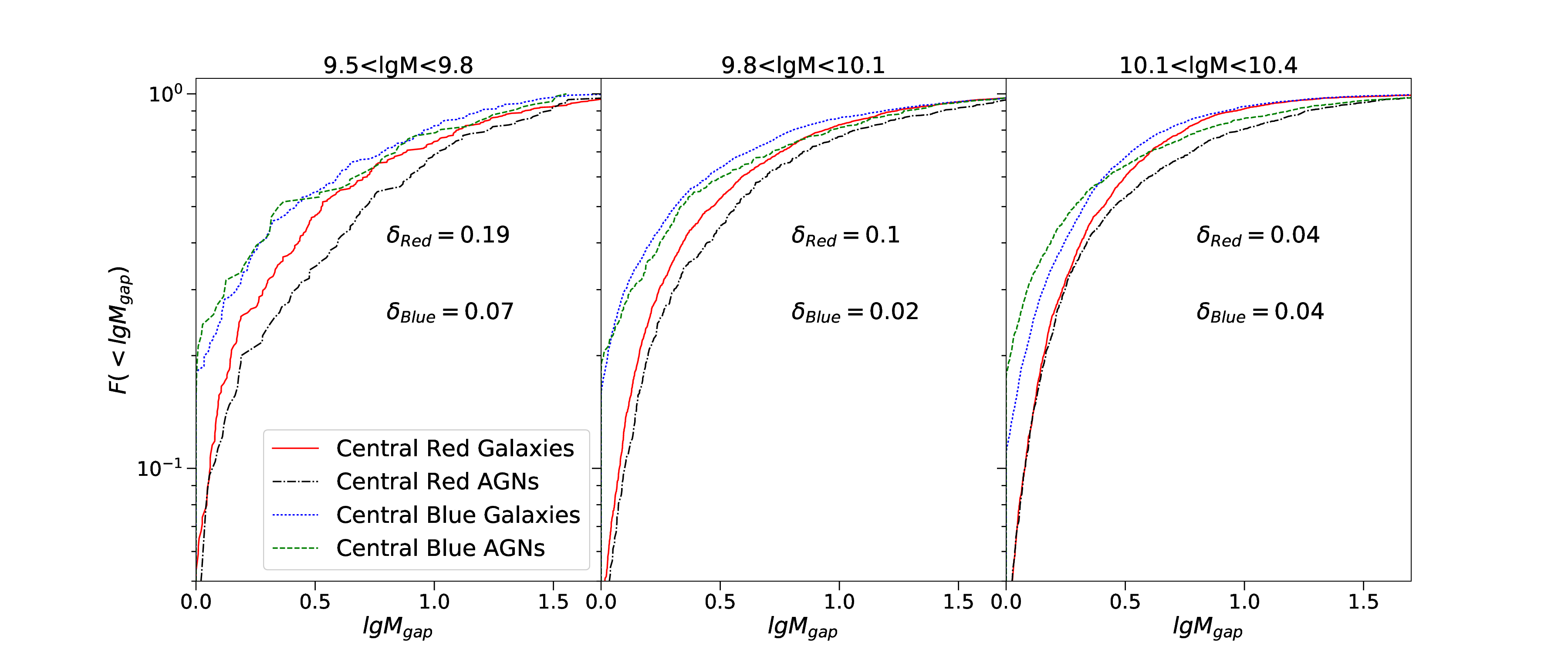,width=\textwidth}
   \end{center}
   \caption{Cumulative fractions of stellar mass gap of galaxy groups,
     $\log_{10}M_{gap}$, are plotted for AGN (dashed lines) and control
     galaxies (solid lines), in different stellar mass bins (indicated
     above each panel) and for red (black/red) and blue (blue/green)
     colors. The difference in the median value of $\log_{10}M_{gap}$
     between each AGN sample and the corresponding control sample is
     indicated.}
   \label{fig:gap}
\end{figure*}

In Figure~\ref{fig:com-agn-gal-halo} we compare the cumulative
distribution of the halo mass for the AGN and control samples in
different mass bins, but for central and satellite galaxies
separately. For this plot we have taken the halo mass estimates, as
well as the central/satellite classification from the SDSS group
catalogue. The upper to lower panels present results for AGN in red
and blue galaxies, respectively. Panels from left to right correspond
to different stellar mass bins. The AGN and control galaxies show
identical halo mass distributions in all cases regardless of color, if
they are central galaxies of their halos. In case of satellite
galaxies, in contrast, AGN are found in less massive halos than
control galaxies, and the effect is seen mainly for the red
populations and is stronger at lower masses. 

In fact, the lower halo mass of the AGN samples of red colors can be
seen from Figs~\ref{fig:agn-stm-color} and~\ref{fig:agn-stm-d4000}.
For the samples with red colors (or low-$D_n(4000)$) and lowest
stellar masses, the AGN antibias is detected out to $\sim10$Mpc, much
larger than the scale of 1-3Mpc in other cases. The clustering
amplitude on scales larger than a few Mpc is known to be an indicator
of dark halo mass. The weaker clustering at large scales thus implies
that the AGN in the low-mass red galaxies are hosted by less massive
halos, compared to the control sample, although the two samples are
already closely matched in many parameters that are known to be
correlated with large-scale clustering. Apparently the clustering
measurements and the halo mass distributions from the group catalog
agree very well with each other.

In Figure~\ref{fig:gap} we compare the stellar mass gap of the host
groups between our AGN and control samples. The stellar mass gap for a
given group, $\log_{10}M_{gap}$, is defined by the difference in 
$\log_{10}M_\ast$ between the most massive galaxy (equivalently the central
galaxy) and the second most massive galaxy (the most massive satellite
galaxy). In this plot we only consider AGN and control galaxies that
are classified as central galaxies of their groups. In addition,  we
restrict ourselves to stellar mass below $10^{10.4}M_\odot$, because
the AGN and control galaxies at higher stellar masses show identical
distributions in this parameter. As can be seen from the figure, in
the case of blue colors, AGN and  control galaxies at given mass tend
to have very similar distributions of $\log_{10}M_{gap}$. Differences
between AGN and control galaxies are seen in the case of red colors
and at stellar mass below $\sim 10^{10}M_\odot$ (the left two panels),
where the host groups of AGN tend to have a larger stellar mass gap
than the host groups of contral galaxies. 

Previous studies have suggested that the luminosity or stellar mass
gap defined this way may be a good tracer of the assembly history of
dark matter halos, and that galaxy groups with a large luminosity gaps
are believed to form earlier than galaxy group with a small luminosity
gap \citep[e.g.][]{Dariush-07, Dariush-10}. A large luminosity or
stellar mass gap has thus been adopted as one of the observational
criteria for identifying ``fossil groups''  \citep[e.g.][]{Jones-03,
  Dariush-07, Dariush-10, Tavasoli-11, Hess-12}. On the other hand, it
has been known for more than a decade that, the clustering of dark
halos is not purely determined by their mass, but also related to
their assembly history
\citep[e.g.][]{Gao-Springel-White-05,Gao-White-07}. Therefore,  the
larger stellar mass gap for the AGN host groups may be suggesting that
AGN prefer to occur in early-formed dark matter halos.

\section{Halo-based modeling of the mass and color dependence of AGN clustering}

In this section we attempt to interpret the co-dependence of AGN
clustering on the stellar mass and color of host galaxies, in the
context of halo occupation models. The clustering difference between
AGN and control samples are seen only at scales between about 100kpc
and 1-3 Mpc, where the AGN are more weakly clustered than the control
galaxies with the same mass, color and structural properties. On
larger scales, the AGN and control galaxies present similar clustering
amplitudes, indicating that they are hosted by halos of similar dark
matter mass. At intermediate scales, as shown in L06, the antibias
observed for the full AGN sample can be simply interpreted if AGN
are preferentially found in central galaxies, which are located at the
center of dark matter halos. This simple model was able to
successfully reproduce the projected 2PCF for both the AGN and the
control sample of non-AGN, if a central fraction of 84\% and 73\% were
adopted for the two samples respectively.

In \S~\ref{sec:clustering}, we showed that the antibias discovered in
L06 is essentially dominated by a subset of AGN that are hosted by
low-mass red galaxies, while the AGN in galaxies of blue colors or
high masses present similar clustering properties to control galaxies.
Previous studies of halo occupation models for galaxy samples selected
by mass (or luminosity) and color have shown that the halo occupations
depend strongly on these two properties \citep[e.g.][]{Zehavi-11}.
Therefore, the mass and color dependence of galaxy clustering must be
considered in any halo-based models before one can further model the
clustering of AGN. In the rest of this section, we will first apply
two commonly-applied halo models, the subhalo abundance matching model
(SHAM) and the subhalo age distribution matching model (SADM), to
assign a stellar mass and a color to each model galaxy in our
simulation. We will then select AGN from the model galaxies by
applying a simliar model to that proposed in L06, which assumes AGN to
be preferentially found at the center of dark halos. 

\subsection{Subhalo abundance matching and subhalo age distribution matching}

Our models are constructed based on the Millennium Simulation  (see
\S~\ref{sec:simulation}), for which dark matter halos and subhalos at
different snapshots are identified, and halo merger trees are
constructed  describing the growth history of the halos. We take all
the subhalos at $z=0$, and assign a stellar mass to each subhalo by
applying the subhalo abundance matching model (SHAM).  This model has
been widely applied for linking galaxies of different stellar masses
to halos of different dark matter masses \citep{Vale-Ostriker-04,
  Conroy-Wechsler-Kravtsov-06, Shankar-06,
  Conroy-Wechsler-Kravtsov-07, Baldry-Glazebrook-Driver-08, Moster-10,
  Guo-10, Neistein-11a, Neistein-11b, Yang-12, Li-12}.  In this model
each (sub)halo is assumed to host a galaxy, of which the stellar mass
is an increasing function of the maximum mass ever attained by its
halo. In practice, the relationship between dark matter halo mass
$M_{200}$ and galaxy stellar mass $M_\ast$ is obtained simply by
matching the number density of halos with mass above $M_{200}$ with
the number density of galaxies with stellar mass above $M_\ast$. To
the end, we have used the stellar mass function of galaxies in the
local  Universe estimated by \citet{Li-White-09} from the SDSS/DR7
galaxy sample,  which is updated in \citet{Guo-10} using the SDSS
``model'' magnitudes instead of ``Petrosian'' magnitudes.

\begin{figure*}
  \begin{center}
    \epsfig{figure=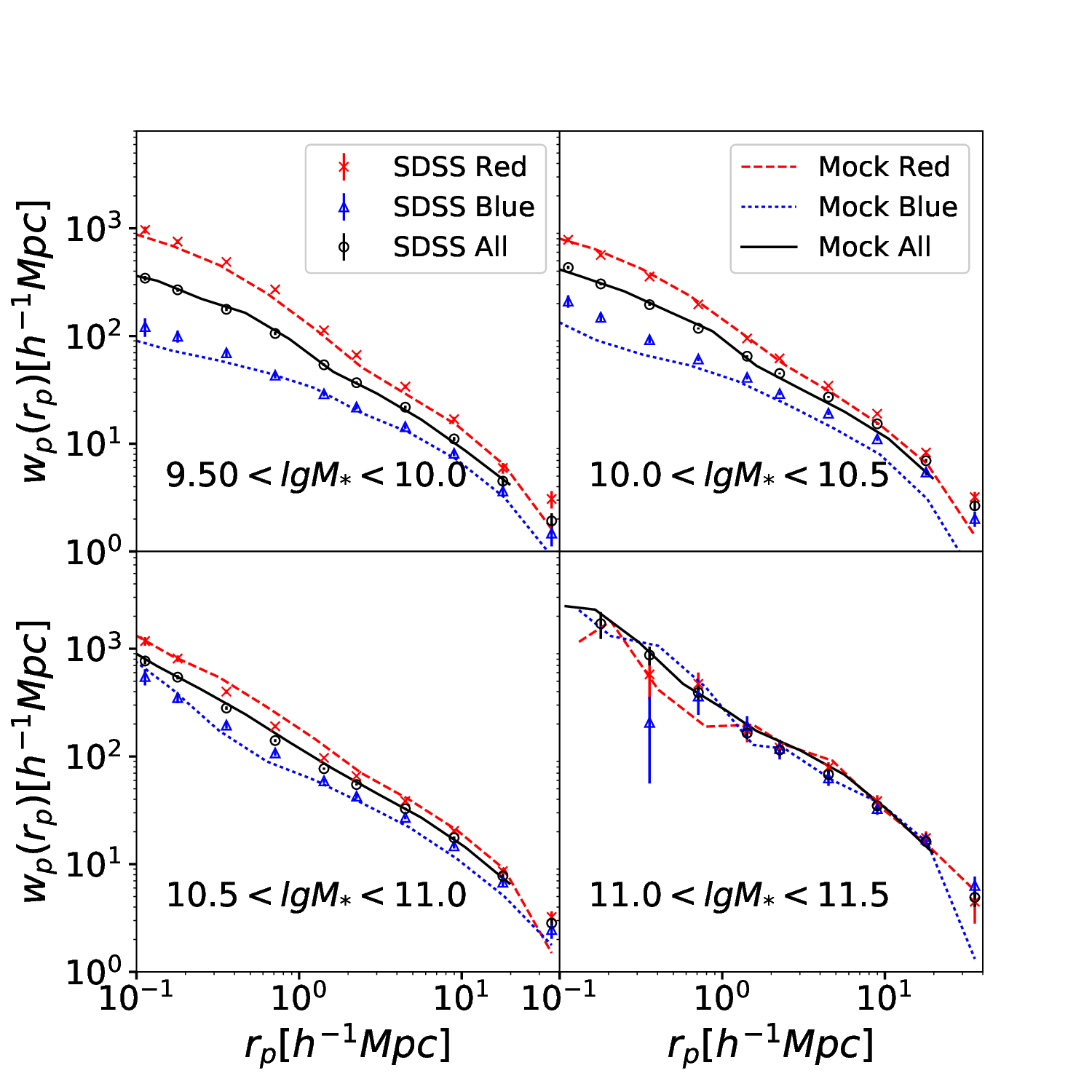,width=0.49\textwidth}
    \epsfig{figure=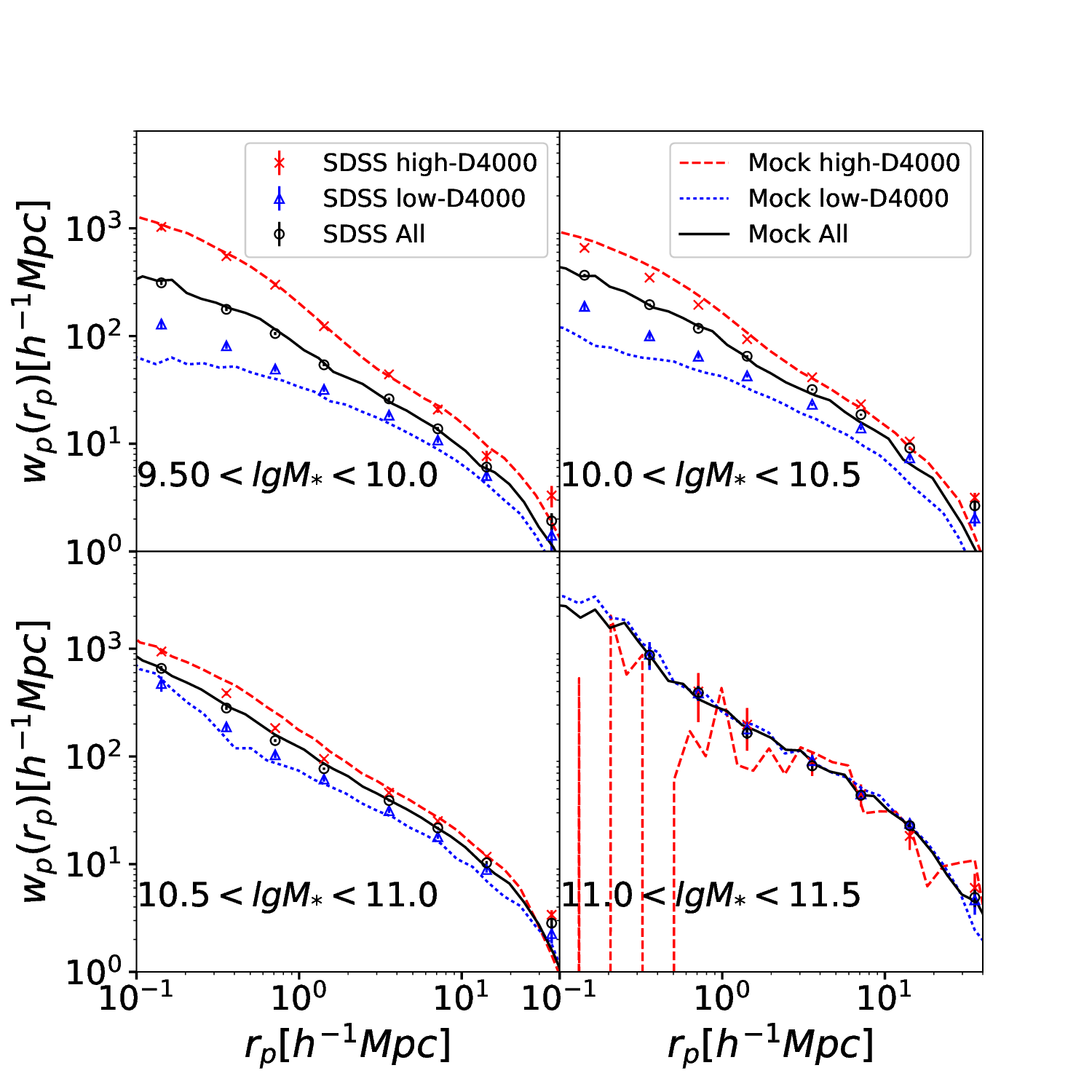,width=0.49\textwidth}
  \end{center}
  \caption{Comparison of projected auto-correlation function of all
    galaxies in the SDSS (symbols) and in our model (lines).
    Different panels are for different stellar mass bins, as
    indicated.  In each panel, the red and blue points/lines represent
    red and blue galaxies, while the black points/line represent all
    the galaxies in the given mass range.}
  \label{figs:all-color}
\end{figure*}

For each model galaxy, we then further assign a color or a
$D_{n}(4000)$  by applying the subhalo age distribution model (SADM),
as recently presented by \citet{Hearin-Watson-13}. Following these
authors,  we define a redshift $z_{starve}$ for each model galaxy by 
\begin{equation}
	z_{starve}=\max(z_{form},z_{acc},z_{char}),
\end{equation}
where $z_{form}$ is the halo formation time at which the dark matter
halo transitions from the fast-accretion regime to the slow-accretion
regime \citep{Wechsler-02}, $z_{acc}$ the epoch when the galaxy was
last the central galaxy of its own halo,  and $z_{char}$ is the epoch
when the halo mass exceeds $10^{11.5}M_{\ast}h^{-1}$.  We note that
the mass threshold adopted here for defining $z_{char}$ is slightly
lower than the value suggested in \citep{Hearin-Watson-13},
$10^{12}M_{\ast}h^{-1}$, as the new mass threshold allows our model to
better reproduce the observed color dependence of galaxy clustering in
this work. As discussed in \citep{Hearin-Watson-13}, the $z_{starve}$
defined this way is expected to encompass physical characteristics of
halo mass assembly that may be responsible for quenching the star
formation in the galaxy, thus a driving parameter for its color or
mean stellar age.  At fixed stellar mass, galaxy color is assumed to
be an increasing  function of $z_{starve}$, and so can be determined
by matching the observed  color distribution of galaxies at the given
stellar mass with the  $z_{starve}$ distribution of dark halos
corresponding to the same stellar mass.

Applications of the SHAM and SADM models to the Millennium Simulation
results in a complete mock catalog of model galaxies at $z=0$,  each
assigned both a stellar mass $M_\ast$ and an optical color $g-r$ (or a
$D_{n}(4000)$). In Figure~\ref{figs:all-color}, we compare the
projected 2PCF $w_p(r_p)$ as measured for the reference galaxy sample
and the model galaxy catalog, for different stellar mass intervals and
for subsets of galaxies with red/blue colors (upper panels), or
high-/low-$D_{n}(4000)$ (lower panels) at fixed mass. We adopt a
mass-dependent color divider to divide galaxies into red and blue
subsets, $g-r = -1.399 + 0.2168 \times \log_{10}M_\ast$,  determined by
fitting a double Gaussian profile to the distribution of $g-r$ at
fixed $M_\ast$ following the method described in  \cite{Li-06b}.  A
mass-dependent divider of $D_{n}(4000)$ is determined in the same way
and applied to divide galaxies at fixed mass into subsets of  high-
and low-$D_{n}(4000)$.

The figure shows that, as expected, the co-dependence of galaxy
clustering on  stellar mass and optical color (or $D_{n}(4000)$)  can
be well reproduced with the model, particularly for the red galaxies
(or those with higher $D_{n}(4000)$) and for the galaxies as a
whole. For blue galaxies or galaxies of low-$D_{n}(4000)$, the  model
works well at all masses except the interval of
$10<\log_{10}(M_\ast/M_\odot)<10.5$, where the model predicts weaker
clustering at all the scales probed.  A similar discrepancy is also
observed in \cite{Hearin-Watson-13}, where the model underpredicts the
clustering  on scales above $\sim1$Mpc for the blue galaxies with
$r$-band absolute  magnitudes in the range $-20<M_r<-21$, which is
roughly corresponding  to the mass range mentioned above, according to
the $r$-band luminosity  function and stellar mass function of SDSS
galaxies \citep[e.g.][]{Blanton-03a, Li-White-09}. We note that, in
the lowest-mass bin ($9.5<\log_{10}(M_\ast/M_\odot)<10$), the clustering of
blue galaxies and those with low $D_{n}(4000)$ in the model is also
slightly lower than the observation for scales smaller than
$\sim1$Mpc. Despite these slight discrepancies, overall the model is
in good agreement with the data, and thus forms a good basis for us to
further model the dependence of AGN clustering on galaxy mass and
color. We will keep these discrepancies in mind when modelling the AGN
clustering in the rest of the paper, though. Given the highly similar
results for $(g-r)$ and $D_{n}(4000)$, as seen in both the previous
section and the current one, we will concentrate on $(g-r)$ in what
follows. We  note that we have done the same analysis for
$D_{n}(4000)$, finding very similar results to $(g-r)$.

\subsection{Modelling the AGN clustering with the simple model of L06}

The mass and color dependence of the central fractions are well
expected by the simple model of L06, in which $f_c$ is the driving
parameter for the antibias of AGN relative to the control sample. In
the previous section we have shown that the antibias is observed only
for the AGN in red galaxies with masses below
$\sim10^{10.5}M_\odot$. Therefore, it is natural to expect that the
same model as proposed in L06 will well explain the $w_p(r_p)$
measurements for the samples of different masses and colors as
measured  in this work. We apply the model of L06 to select AGN from
the model galaxy catalog constructed above, according to the
mass-dependent central fraction of AGN and control galaxies, but for
red and blue colors separately. We divide the model galaxies into  red
and blue populations according to the $(g-r)$, in the same was as done
for the real sample. For a selected sample of model AGN, we construct
a control sample by requiring it closely match the model AGN sample in
both stellar mass and color. We don't consider other parameters such
as concentration and stellar velocity dispersion which are not
available in our current model, but we argue that the dependence of
clustering on those parameters are known to be much weaker than the
dependence on mass and color.

\begin{figure*}
  \begin{center}
    \epsfig{figure=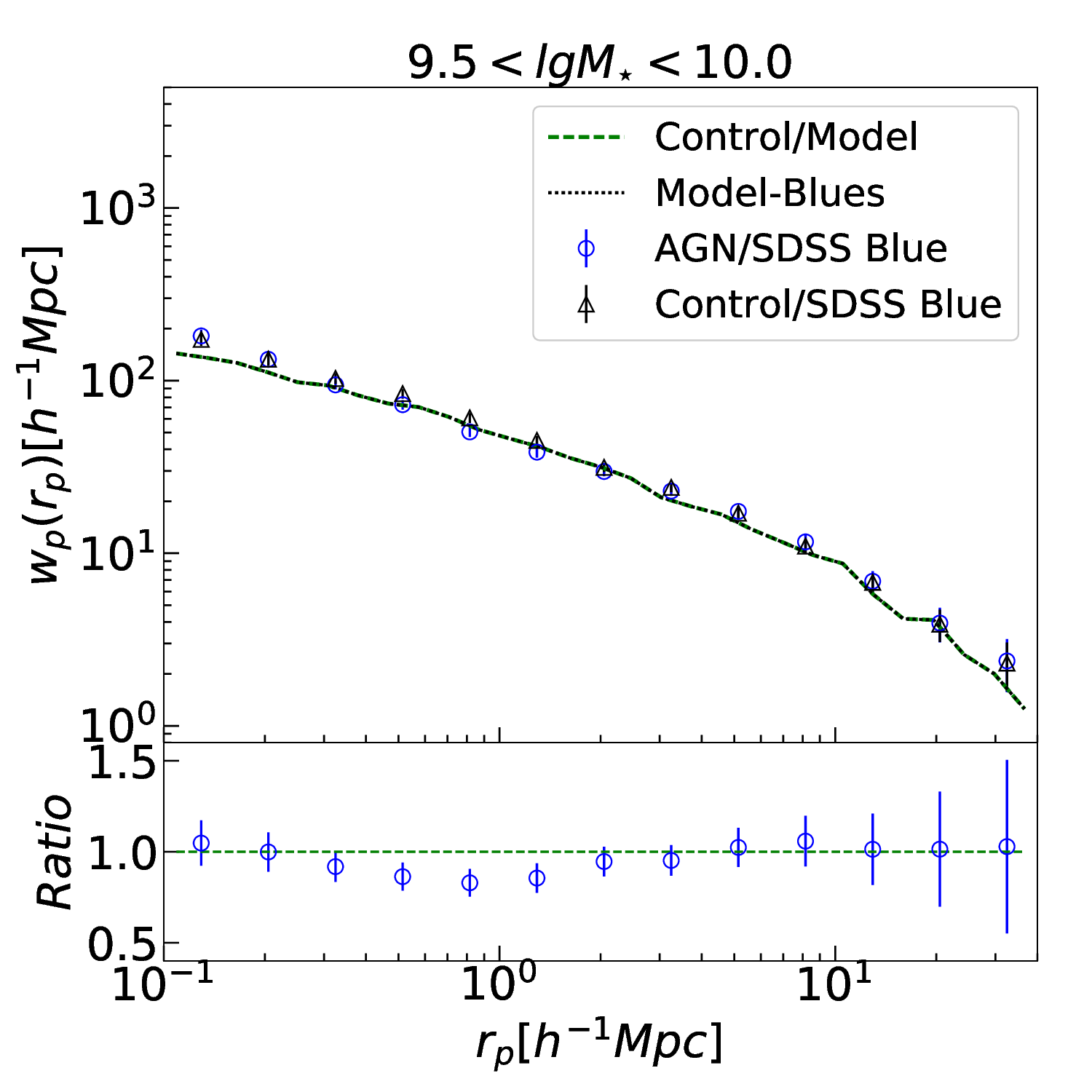,width=0.33\textwidth}
    \epsfig{figure=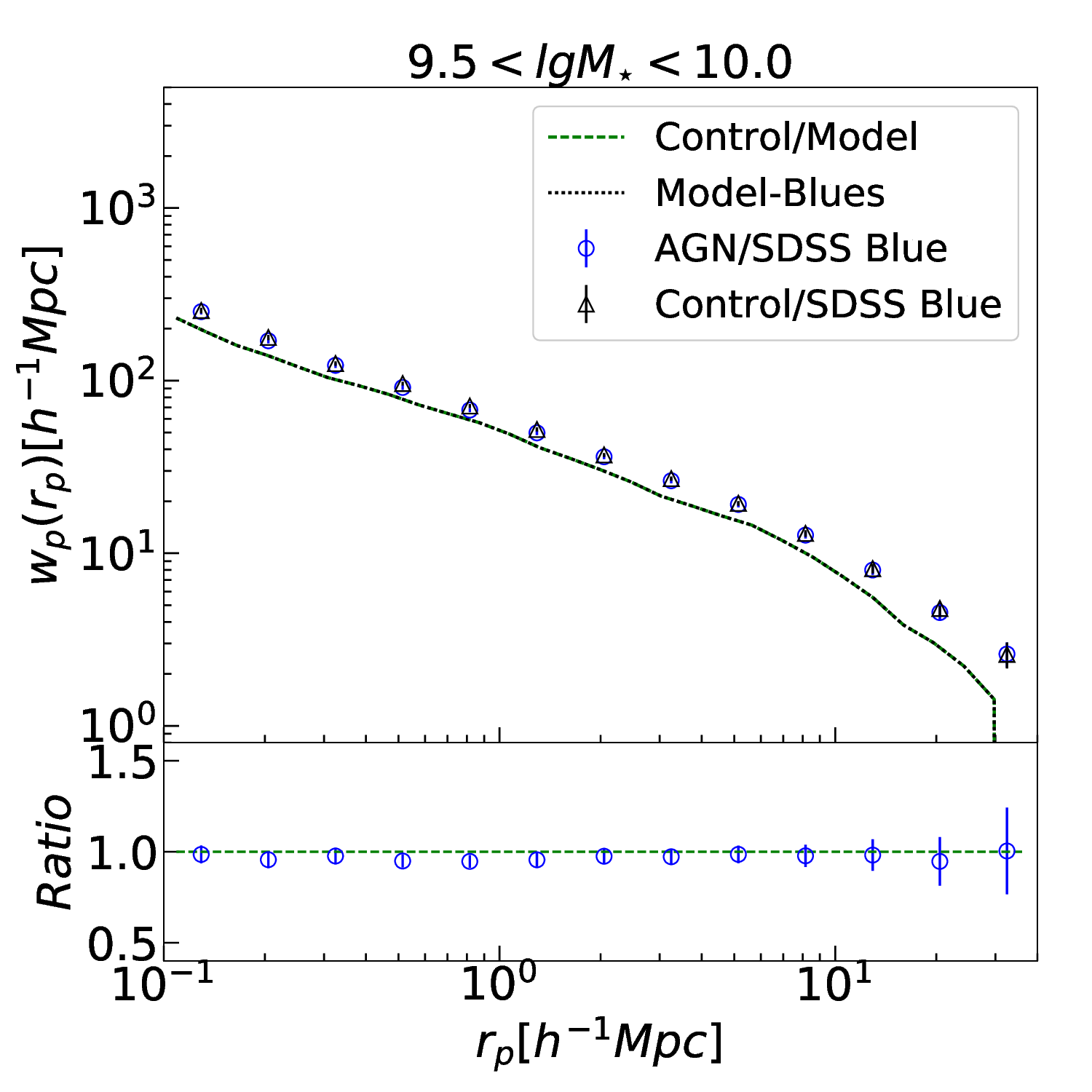,width=0.33\textwidth}
    \epsfig{figure=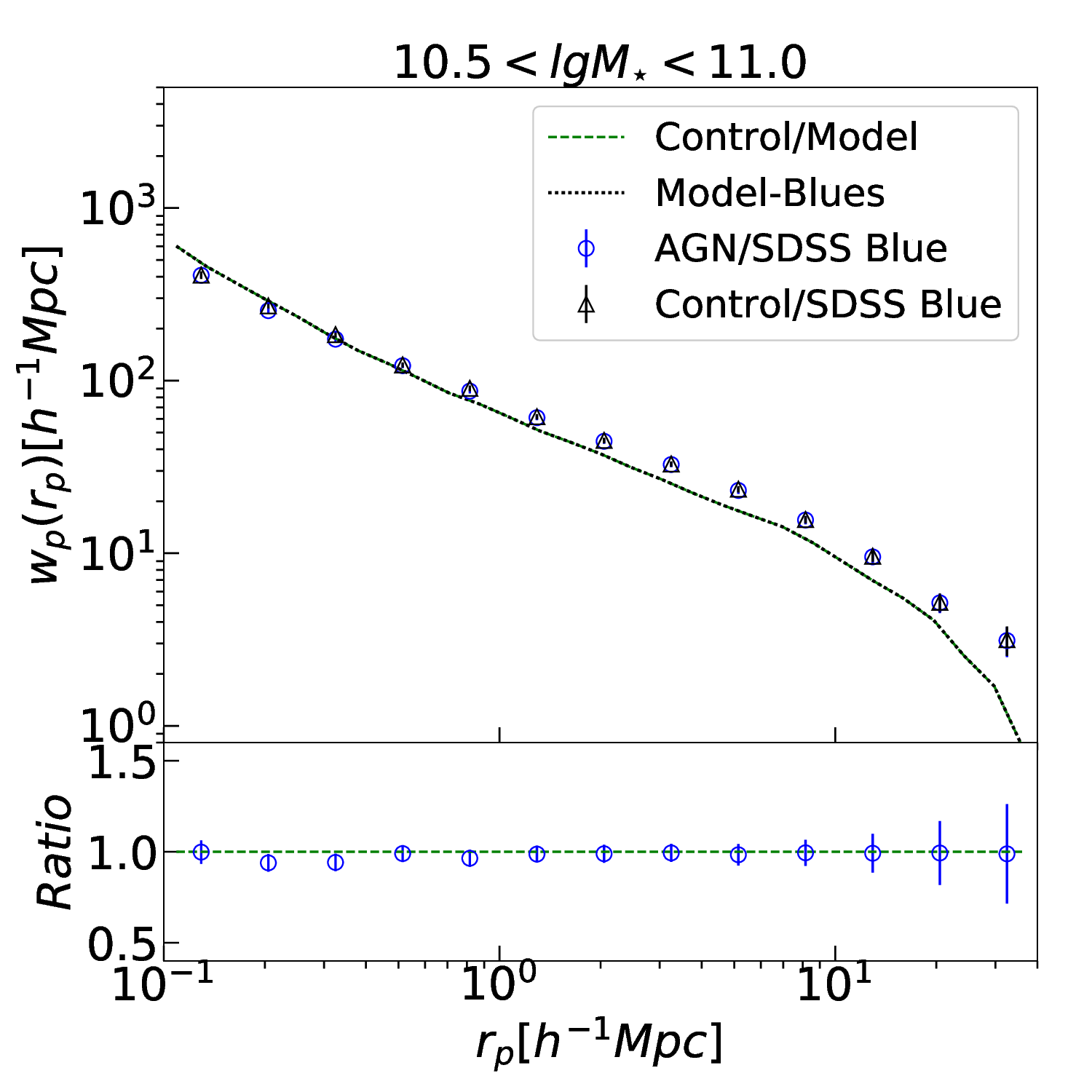,width=0.33\textwidth}
  \end{center}
  \caption{Top: open circles and triangles show the $w_p(r_p)$ for AGN
    and control galaxies with blue colors, while the green dashed and
    black dotted lines show the results for the corresponding control
    samples. Different panels are for different stellar mass bins as
    indcated. Bottom: ratio of $w_p(r_p)$ between AGN and control
    galaxies.  Symbols/lines are the same as in the top panel.}
  \label{fig:com-obs-model-AGN-blue}
\end{figure*}

\begin{figure*}
  \begin{center}
    \epsfig{figure=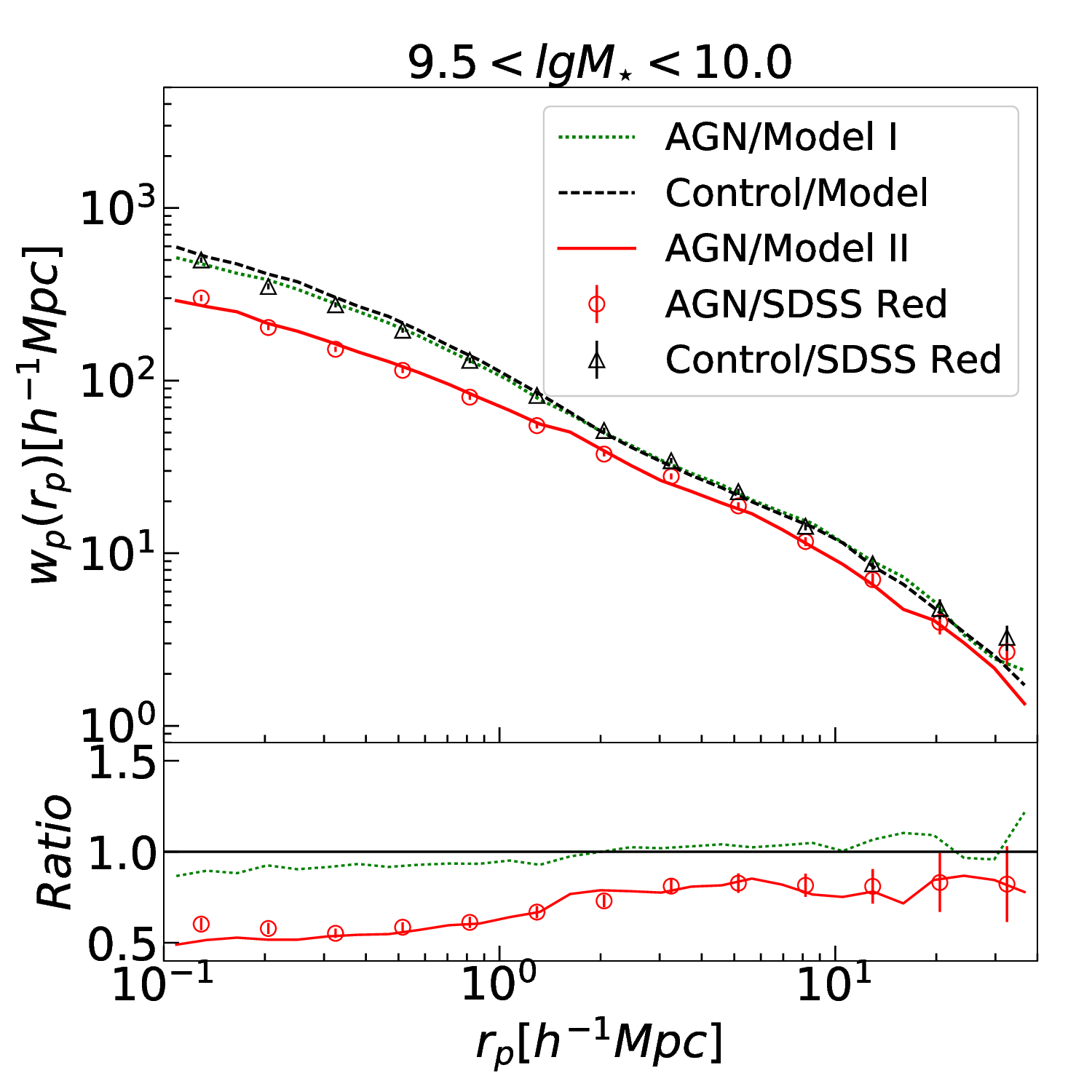,width=0.33\textwidth}
    \epsfig{figure=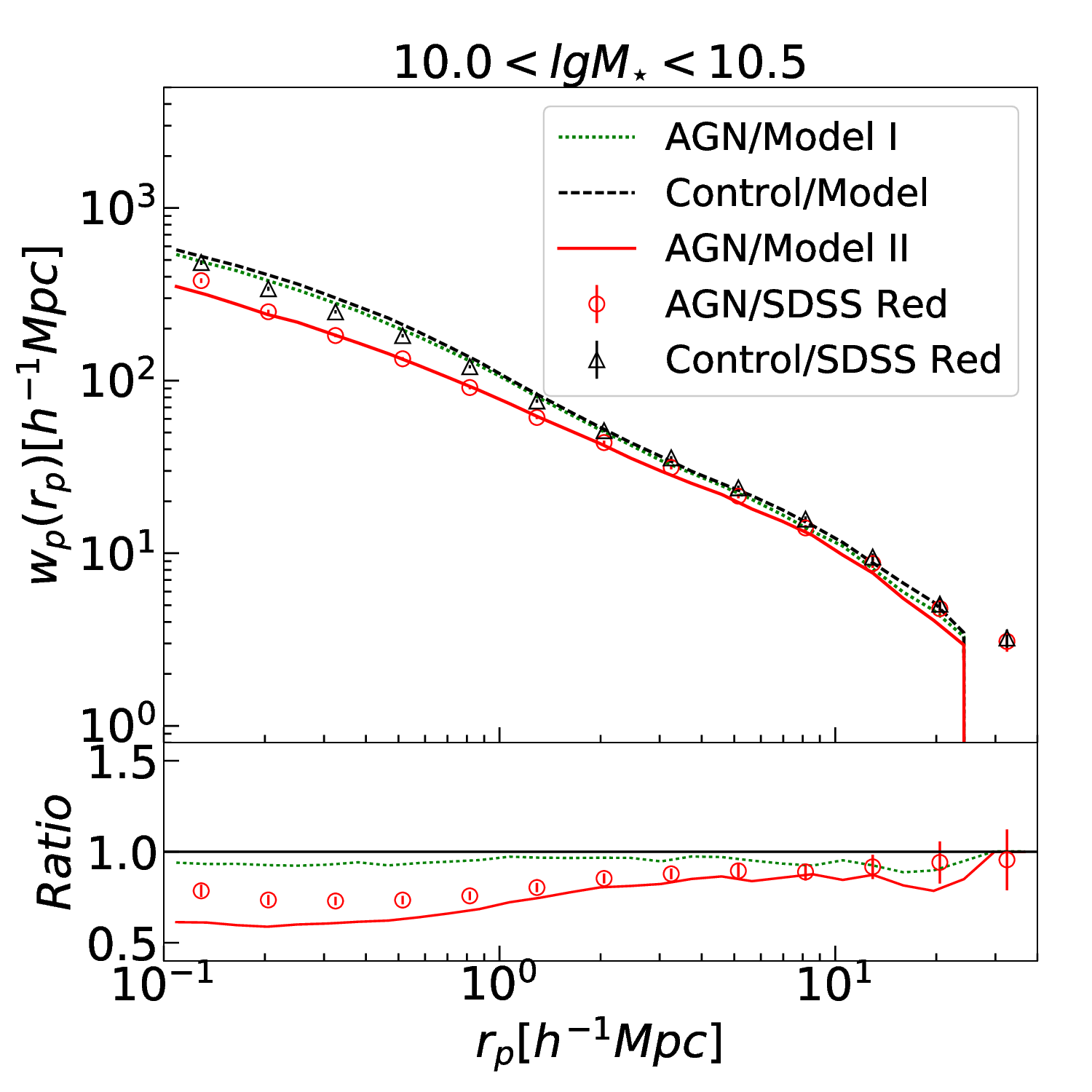,width=0.33\textwidth}
    \epsfig{figure=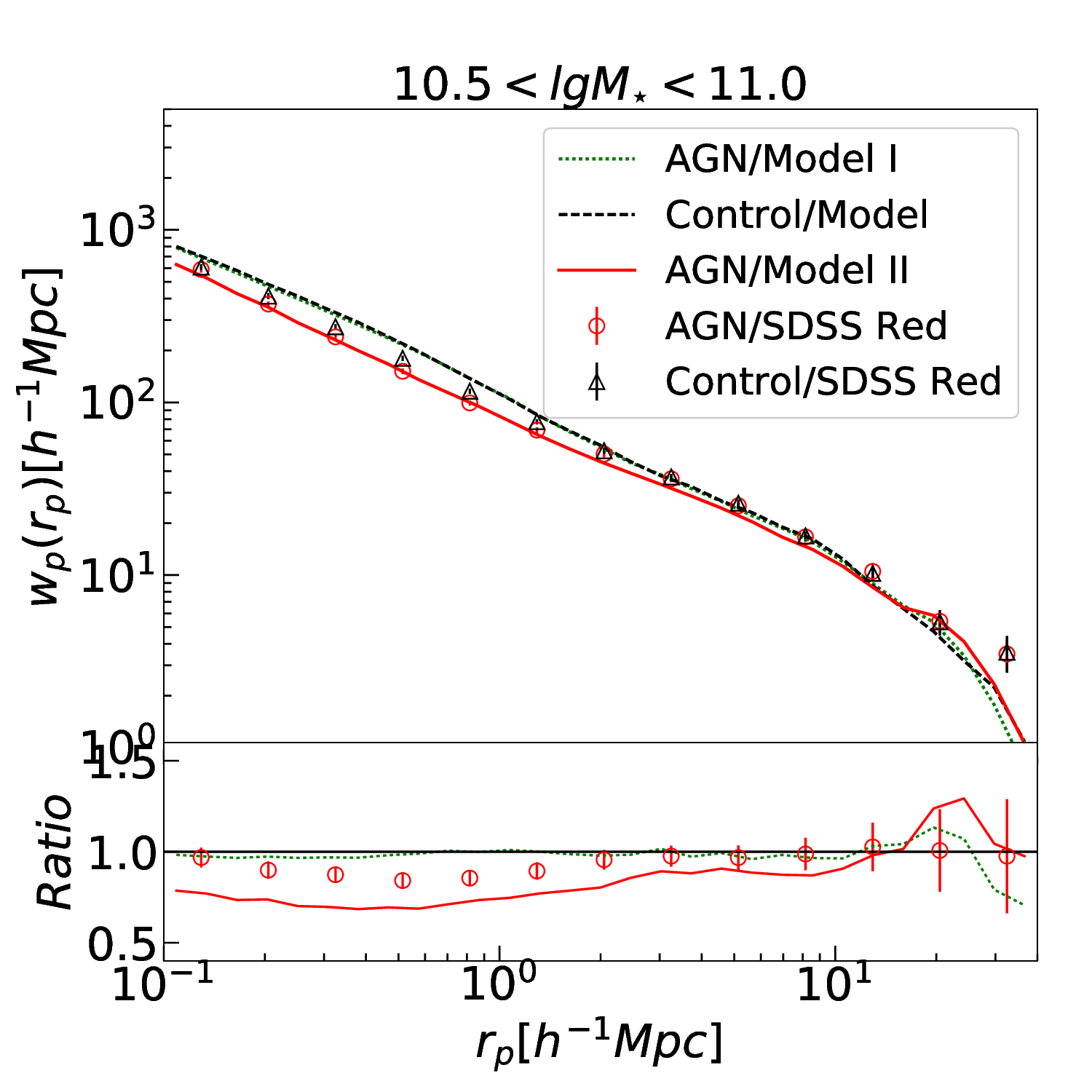,width=0.33\textwidth}
  \end{center}
  \caption{Top: open circles and triangles show the $w_p(r_p)$ for AGN
    and control galaxies with red colors. The three lines are  the
    results for the models. The blue dashed line for the  control
    galaxies in the models, while the green dotted and  solid red
    lines are for the AGN in the two models separately (see the text
    for details. Bottom: ratio of $w_p(r_p)$ between AGN and control
    galaxies.  Symbols/lines are the same as in the top panel.}
  \label{fig:com-obs-model-AGN-red}
\end{figure*}

For the AGN/controls of blue colors, since AGN and control galaxies
present similar central fractions at fixed mass, we simply select a
random subset of blue galaxies to be our model AGN, thus essentially
assuming AGN are found equally in every galaxy if its color is
blue. Figure~\ref{fig:com-obs-model-AGN-blue} compares  the $w_p(r_p)$
predicted by this model with the observation, for the three stellar
mass intervals. As expected, the AGN and control galaxies in the model
show identifical clustering properties at given mass, which well match
the $w_p(r_p)$ of the SDSS AGN and control samples. We note that the
model slightly underpredicts the clustering amplitude at scales above
a few Mpc for the two high-mass bins, which is the problem of the
model of the blue galaxy population as pointed out in the previous
subsection. For AGN in blue galaxies, it is clear that they  show no
preference in terms of halo mass and environment of all scales, when
compared to control galaxies of the same mass and color.

For the AGN/controls of red colors, we select AGN from the red
galaxies in the model catalog, requiring the relative difference of
$f_c$ between the AGN and control sample to follow the observation
(see Figure~\ref{figs:agncolorfra}), which can be described by
\begin{equation}\label{eqn:fcentral}
  \delta_{f_c}/f_{c,control} = 0.1434\times (\log_{10}M_\ast)^{2} 
    -3.1092\times \log_{10}M_\ast +16.8724.
\end{equation}
The $w_p(r_p)$ predicted by this model are compared to the
observation in Figure~\ref{fig:com-obs-model-AGN-red}, again for the
three mass intervals  separately. In the figure, the green dotted
lines are for the AGN in the model, while the black dashed lines are
for the control samples of model galaxies. The green dotted lines in
the smaller panels present the AGN-to-control ratio of $w_p(r_p)$ for
the model. To our surprise, the  AGN and control galaxies in the model
show similar clustering amplitudes on all scales and at all masses,
with no significant antibias everywhere.  The model indeed predicts
slightly weaker clustering on scales below a few Mpc for AGN in the
two low-mass bins, but the difference  is much less significant than
the observed antibias.

\begin{figure}
  \begin{center}
    \epsfig{figure=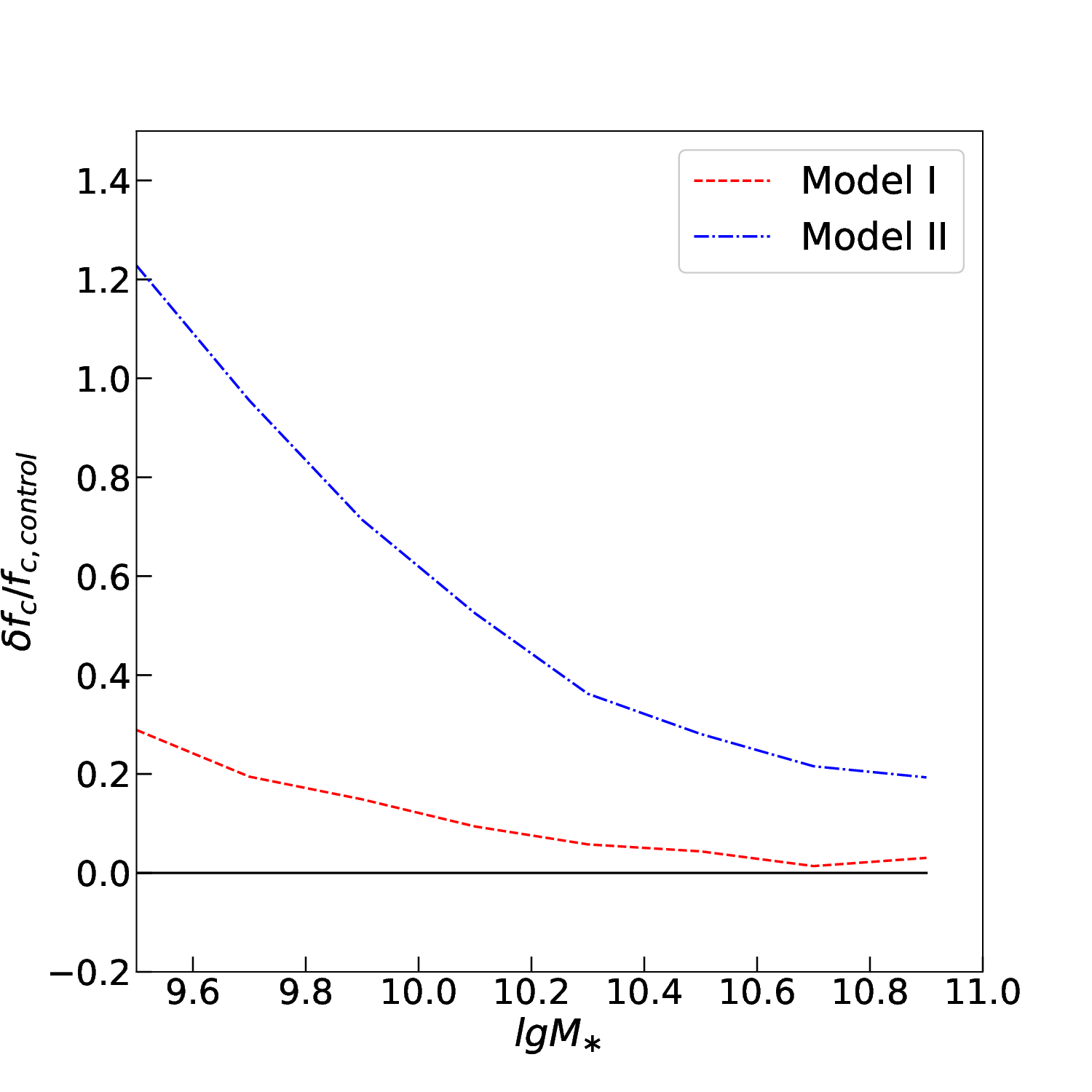,width=0.49\textwidth}
  \end{center}
  \caption{Fraction of AGN in central galaxies as a function of stellar
    mass, as adopted by the two models separately. See the text for
    details.}
  \label{fig:com-diff-model-cenfrac}
\end{figure}

\begin{figure}
  \begin{center}
    \epsfig{figure=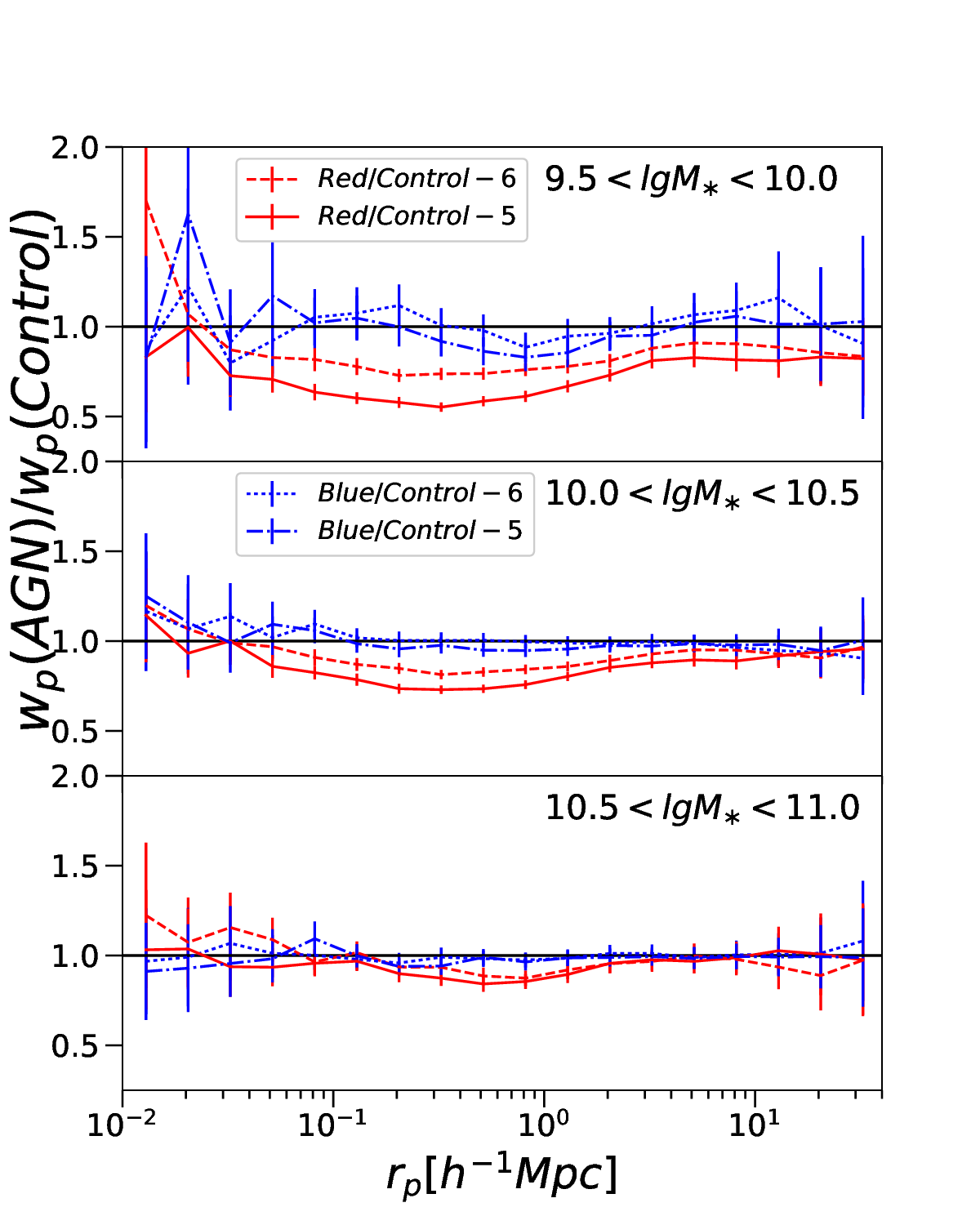,width=0.49\textwidth}
  \end{center}
  \caption{AGN-to-control ratio of $w_p(r_p)$ for different stellar
    mass bins (indicated in each panel), and for AGN in red (red
    lines) and blue (blue lines) galaxies separately. In each panel,
    the dotted lines are for the AGN and control samples matched in
    five parameters, while the dashed lines for the samples matched
    additionally in central fraction.}
  \label{fig:com-agn-control-group}
\end{figure}

L06 has shown that a higher central fraction for AGN can make the
model reproduce the observed antibias at scales between $\sim100$kpc
and a few Mpc. In that work, however, the central fraction was assumed
to be a free parameter and constrained purely by the clustering
measurements. It is interesting to see whether the model can still
work if $f_c$ is a free parameter. In this case, $f_c$ for AGN in a
given mass bin is determined by fitting the $w_p(r_p)$ of the model to
the observed one. The clustering measurements of the AGN sample in the
best-fit model are shown also in
Figure~\ref{fig:com-obs-model-AGN-red}, but in solid red
lines. Results of the control samples remain unchanged, and so are not
plotted. As can be seen, the model can truely reproduce both the
$w_p(r_p)$  measurements and the AGN-to-control antibias at scales
below a few Mpc, although the antibias in the model appears to be even
stronger. For convenience, in what follows we will call this model as
{\tt Model II} and the model which adopts the observed $f_c$ as {\tt
  Model I}.

In Figure~\ref{fig:com-diff-model-cenfrac} we plot $f_c$ as a function
of stellar mass for both Model II (the blue dotted line) and Model I
(the red dashed line). The central fractions of AGN in Model II are
higher by a factor of $\sim$4 than the fractions  estimated from the
SDSS group catalog. We should point out that, for simplicity, the
fitting for Model II is done for the three mass bins simultaneously,
by multiplying both sides of Eqn.~(\ref{eqn:fcentral}) by a common
factor.  We can well expect the model to better match the data in the
two high-mass bins if the fitting is done for the different mass bins
independently. Our purpose here is not to obtain an accurate
model. Rather, we aim at demonstrating that the central fraction alone
is able to explain the observed mass-dependence of  the AGN antibias,
and our result here shows this is indeed the case.  

On the other hand, however, the fact that the model requires central
fractions  much higher than the real sample implies that the observed
antibias of AGN cannot be purely explained by a higher $f_c$.  To test
this out, we have done an additional analysis of the real sample, in
which we further require each control sample to have the same central
fraction as the corresponding AGN sample.  Results of this analysis
are shown in Figure~\ref{fig:com-agn-control-group}, where we plot the
AGN-to-control $w_p(r_p)$ ratio as a function of $r_p$. The three
panels are for the different mass intervals, and in each panel the
red/blue lines are for the red and blue subsamples respectively. The
results of the new analysis are shown in dashed lines, and the results
from the previous section where the control samples are not matched in
central fraction are shown in dotted lines, for comparison. Generally,
the AGN antibias becomes weaker, but is still significantly detected,
as can be seen in the low-mass red subsamples. This result clearly
demonstrates that the central fraction can partly but not entirely
explain the AGN antibias.

\section{Summary and Conclusions}
\label{sec:summary}

We have studied the joint dependence of AGN clustering on  galaxy
stellar mass and color (or $D_n(4000)$), using a sample of
$\sim10^{5}$ narrow-line AGN and a sample of about half a million
reference galaxies, both selected from the final data release of the
Sloan Digital Sky Survey \citep[SDSS/DR7][]{Abazajian-09}. The AGN are
divided into different stellar mass bins, and for a given mass bin
they are further divided into red and blue subsamples according to the
optical color $(g-r)$ or $D_n(4000)$ of their host galaxies. For each
AGN sample, we have constructed a control sample from all the
reference galaxies which are closely matched with the AGN sample in
redshift, stellar mass, color, stellar velocity dispersion and
concentration index. For both the AGN sample and its corresponding
control sample, we then estimate the projected cross-correlation
function $w_p(r_p)$ with respect to the reference galaxy sample, and
compare the $w_p(r_p)$ measurements between AGN and control
galaxies. Next, we make use of the SDSS/DR7 galaxy group catalogue
constructed by \cite{Yang-07} to further examine the halo properties
of the AGN and control galaxies, but for the AGN/controls in central
and satellite galaxies separately.  We consider three properties of
the host groups: the fraction of central galaxies in the sample
($f_c$), the halo mass distribution,  and the stellar mass gap between
the two most massive galaxies of a given group. Finally, we apply the
commonly-adopted subhalo abundance matching model and subhalo age
distribution matching model to populate the (sub)halos in the
Millennium Simulation with galaxies of different stellar masses and
colors, from which we further select AGN by applying the simple model
of \cite{Li-06a} (L06) which assumes AGN to be found more
preferentially at the center of dark matter halos. 

We have reproduced the AGN antibias as originally found in L06, that
is, at scales between about 100kpc and a few Mpc the AGN as a whole
are more weakly clustered than the carefully-matched control
sample. When we divide the AGN and control galaxies into subsamples by
stellar mass and color (or $D_n(4000)$), we obtain  the following
results.

\begin{itemize}
\item The antibias previously observed from the full AGN sample is
  hold only when the AGN host galaxies are less massive than
  $M_\ast\sim10^{10.5}M_\odot$ and have red colors (or high
  $D_n(4000)$), while AGN hosted by red galaxies of higher masses or
  blue galaxies of all masses show almost identical clustering
  properties to the control galaxies at all scales. This result is
  shown to be independent of black hole mass, AGN power and AGN type.
\item AGN in blue galaxies are found to have similar mass-dependent
  central galaxy fraction to the control galaxies. The AGN and control
  galaxies in this case also show similar halo mass distributions, and
  this is true even when the central AGN and satellite AGN are
  considered separately.
\item AGN in red galaxies have higher central fraction than the
  control galaxies, with larger difference at lower stellar masses.
  On average, the host groups of the AGN associated with red satellite
  galaxies  appear to have lower-than-average dark matter masses,
  while the host groups of the AGN associated with red central
  galaxies tend to have larger stellar mass gap, indicative of earlier
  formation time of their dark halos. 
\item A simple halo-based model in which the AGN are preferentially
  found in central galaxies can in principle reproduce the mass and
  color dependence of the AGN clustering in the SDSS sample. However,
  the central fraction of AGN in the best-fit model is a factor of
  $\sim$4 higher than the central fraction of the real AGN sample,
  implying that the central fraction alone cannot fully explain the
  AGN clustering properties.
\end{itemize}

The strong mass and color dependence of the AGN clustering is
striking.  If hosted in blue galaxies, the AGN show the same
clustering and the same group/halo properties as the  control galaxies
of similar mass, color and structural parameters. The same conclusion
is also true for AGN in massive galaxies with
$M_\ast\ga10^{10.5}M_\odot$. In other words, the AGN activity in blue
galaxies or massive galaxies  is regulated only by the physical
processes internal to galaxies, with no correlations with environment
on scales larger than the size of individual galaxies.  Previous
studies of optically-selected AGN have well established that more than
a half of the AGN population in the local Universe are found in
massive galaxies with stellar mass above $10^{10.5}M_\odot$
\citep[e.g.][]{Heckman-80, Ho-97, Kauffmann-03c, Hao-05}. Therefore,
the majority of the  AGN activity in the local Universe is not driven
by environmental effects.

The unbiased clustering of AGN in blue galaxies is also a very
interesting result. Studies of the correlation of local black hole
growth with host galaxy properties have demonstrated that black holes
grow more rapidly in galaxies with younger stellar populations
\citep{Kauffmann-03c,CidFernandes-04}. Furthermore,
\cite{Kauffmann-Heckman-09} have revealed two distinct regimes of
black hole growth in low-redshift galaxies, which are strongly linked
to the star formation history of the central stellar population of the
host galaxies. In one regime, where the host galaxies are undergoing
significant central star formation, the distribution of the Eddington
ratio ($L_{AGN}/_{Edd}$) universally follows a log-normal form,  which
is independent of both the black hole mass and current star formation
rate. In the other regime, where the host galaxies have little or no
ongoing star formation in the central region, the Eddington ratio
presents a power-law distribution with a normalization depending on
the age of the stellar population. 
In a more recent work, \cite{Aird-18} used NIR and X-ray deep imaging
to measure the probability distribution function of AGN accretion
rates as a function of stellar mass for both star-forming
and quiescent galaxy populations. The authors also identified two
different modes of black hole growth in the two types of galaxies,
although their distribution functions of star-forming galaxies
are broader than the lognomal distribution as originally
identified by \citet{Kauffmann-Heckman-09} and agree better with
the Schechter function proposed by \citet{Jones-16}.
Although it is not immediately
clear whether and how the dichotomy in black hole growth is related to
the color-dependent AGN clustering, the two observational results are
interestingly similar in several aspects, e.g. the universal growth
law of black holes with young stellar populations versus the unbiased
clustering of  AGN with blue colors or low $D_n(4000)$, and the
non-universal growth law of black holes with old stellar populations
versus the antibias of AGN with red colors or high $D_n(4000)$.  
In future works it would be interesting to explore possible physical
reasons behind this similarity.

The SDSS group catalog has confirmed the conjecture in L06 that AGN
are found preferentially in central galaxies, and the simple
halo-based model proposed in the same paper can in principle reproduce
the mass and color dependence of the AGN clustering, as shown in this
paper.  However, the best-fit model requires a central fraction which
is substantially too high when compared to the central fraction of the
AGN in the group catalog. This demonstrates that, in addition to the
central fraction, one would need additional factors in order to  have
a complete understanding of the AGN clustering. The group catalogue
provides some clues as we discussed. First, the AGN in red satellites
tend to have lower dark matter halo mass which lead to a weaker
clustering amplitude at scales larger than a few Mpc. Second, the
groups hosting the AGN of red central galaxies tend to form earlier
than the groups hosting the control galaxies of the same mass, color
and structural parameters. This implies that the assembly history of
the host dark halos should play some role in triggering the AGN
activity. 

In a recent study of radio-loud AGN in fossil groups of galaxies,
\citet{Hess-12} found that two thirds of the 30 fossil group
candidates contain a radio-loud AGN at the center of their dominant
elliptical galaxy, which is a large fraction as fossil groups were
believed to be old, quiescently evolving galaxy systems. These authors
argued that the radio luminosity is related to the properties of the
group/cluster environment, as well as the mass assembly of the
dominant ellipgical galaxy. Obviously our finding above is  well
consistent with their work. In fact, the fact that radio galaxies
prefer to reside at the center of elliptically dominant groups have
been noticed in earlier studies \citep[e.g.][]{Best-04,Croston-05}.
These studies suggested that current AGN activity in fossil groups is
linked to the heating of their intergalactic medium (IGM).  On the
other hand, however, some authors suggested that  galaxies with red
colors or old stellar populations may have a reservoir  of cold gas
which can also fuel the central black hole
\citep[e.g.][]{Kauffmann-07, vonderLinden-10}.  The centers of
groups/clusters could well be a preferred environment for  these
galaxies for the high cooling efficiency and cold gas density in the
halo center. 

More studies, both theoretical and observational, are needed in order
to better understand the role of halo assembly history in driving the
AGN activity. Possible ways to go include improving the simple halo
model of L06 by considering the halo formation time as an  additional
parameter to the central fraction, and examining the current
hydrodymatical simulations of galaxy formation with different recipes
on AGN feedback  effects. In addition, next-generation large
spectroscopic surveys of high-redshift galaxies to be carried out in
the next few years will allow us to extend the study of AGN clustering
to redshifts $z>1$.  It would be interesting to see whether the mass-
and color-dependent AGN clustering is already at place at $z\ga1$, an
epoch at which the cosmic density of both black hole accretion rate
and star formation rate are peaked.

\section*{Acknowledgement}

We're grateful to Houjun Mo for helpful discussions.  
We acknowledge the support by National Key Basic Research Program of
China (No. 2015CB857004), National Key R\&D Program of China 
(No. 2018YFA0404502), and the NSFC (No. 11173045, 11233005, 11325314, 
11320101002). The Millennium Run
simulation used in this paper was carried out by the Virgo
Supercomputing Consortium at the Computing Centre of the Max-Planck
Society in Garching. 

Funding for  the SDSS and SDSS-II  has been provided by  the Alfred P.
Sloan Foundation, the Participating Institutions, the National Science
Foundation, the  U.S.  Department of Energy,  the National Aeronautics
and Space Administration, the  Japanese Monbukagakusho, the Max Planck
Society,  and the Higher  Education Funding  Council for  England. The
SDSS Web  Site is  http://www.sdss.org/.  The SDSS  is managed  by the
Astrophysical    Research    Consortium    for    the    Participating
Institutions. The  Participating Institutions are  the American Museum
of  Natural History,  Astrophysical Institute  Potsdam,  University of
Basel,  University  of  Cambridge,  Case Western  Reserve  University,
University of Chicago, Drexel  University, Fermilab, the Institute for
Advanced   Study,  the  Japan   Participation  Group,   Johns  Hopkins
University, the  Joint Institute  for Nuclear Astrophysics,  the Kavli
Institute  for   Particle  Astrophysics  and   Cosmology,  the  Korean
Scientist Group, the Chinese  Academy of Sciences (LAMOST), Los Alamos
National  Laboratory, the  Max-Planck-Institute for  Astronomy (MPIA),
the  Max-Planck-Institute  for Astrophysics  (MPA),  New Mexico  State
University,   Ohio  State   University,   University  of   Pittsburgh,
University  of  Portsmouth, Princeton  University,  the United  States
Naval Observatory, and the University of Washington.

\bibliography{reference}

\label{lastpage}
\end{document}